\documentclass[10pt,journal,compsoc]{IEEEtran}

\usepackage{xcolor}
\usepackage{eso-pic}
\usepackage[nocompress]{cite}
\usepackage{url}
\usepackage{amsmath,amssymb,graphicx}
\usepackage{algorithmic}
\usepackage{algorithm}

\DeclareMathOperator*{\argmin}{arg\,min}
\usepackage{array}
\usepackage{subcaption}

\usepackage{makecell}
\usepackage{diagbox}

\title{Tomographic Sparse View Selection using the View Covariance Loss}

\author{Jingsong~Lin,~\IEEEmembership{Student Member,~IEEE,}
        Amirkoushyar Ziabari,~\IEEEmembership{Senior Member,~IEEE,}
        Singanallur V. Venkatakrishnan,~\IEEEmembership{Senior Member,~IEEE,}
        Obaidullah Rahman,
        Gregery T. Buzzard,~\IEEEmembership{Senior Member,~IEEE,}
        Charles A. Bouman,~\IEEEmembership{Life~Fellow,~IEEE}%
\IEEEcompsocitemizethanks{
\IEEEcompsocthanksitem Jingsong Lin is with the School of Electrical and Computer Engineering, Purdue University.
\IEEEcompsocthanksitem Amir Koushyar Ziabari, Singanallur V. Venkatakrishnan, and Obaidullah Rahman are with Oak Ridge National Laboratory.
\IEEEcompsocthanksitem Gregery T. Buzzard is with the Department of Mathematics, Purdue University.
\IEEEcompsocthanksitem Charles A. Bouman is with the School of Electrical and Computer Engineering, Purdue University.
}
}

\begin{document}
\AddToShipoutPictureFG*{\AtPageUpperLeft{\put(\LenToUnit{0.6in},\LenToUnit{-0.42in}){\parbox[t]{\dimexpr\paperwidth-1.2in\relax}{\fontsize{6.5}{7.5}\selectfont\centering \copyright~2025 IEEE. Personal use of this material is permitted. Permission from IEEE must be obtained for all other uses, in any current or future media, including reprinting/republishing this material for advertising or promotional purposes, creating new collective works, for resale or redistribution to servers or lists, or reuse of any copyrighted component of this work in other works. Accepted version of the article published in IEEE Transactions on Pattern Analysis and Machine Intelligence, vol.~48, no.~10, pp.~11335--11345, Oct.~2026. DOI: 10.1109/TPAMI.2025.3600072}}}}

\IEEEtitleabstractindextext{%
\begin{abstract}
Standard computed tomography (CT) reconstruction algorithms such as filtered back projection (FBP) and Feldkamp-Davis-Kress (FDK) require many views for producing high-quality reconstructions, which can slow image acquisition and increase cost in non-destructive evaluation (NDE) applications.
Over the past 20 years, a variety of methods have been developed for computing high-quality CT reconstructions from sparse views.
However, the problem of how to select the best views for CT reconstruction remains open.
In this paper, we present a novel view covariance loss (VCL) function that measures the joint information of a set of views by approximating the normalized mean squared error (NMSE) of the reconstruction.
We present fast algorithms for computing the VCL along with an algorithm for selecting a subset of views that approximately minimizes its value.
Our experiments on simulated and measured data indicate that for a fixed number of views our proposed view covariance loss selection (VCLS) algorithm results in reconstructions with lower NRMSE, fewer artifacts, and greater accuracy than current alternative approaches.
\end{abstract}

}

\maketitle

\IEEEraisesectionheading{
  \section{Introduction}\label{sec:introduction}
}

\IEEEPARstart{C}{omputed} tomography (CT) is a widely used imaging tool in industrial non-destructive characterization.
It enables visualization of the internal structure of scanned objects, aiding in the understanding of internal features, metrology and the detection of defects and anomalies\cite{iCT_review}.
A typical CT scan collects projections from different orientations around the object, also referred to as \textit{views}, and a reconstruction algorithm then processes these projections to reconstruct an image or volume.
Tuy's condition \cite{Tuy1,Tuy2} provides a general theory to determine the set of views required for perfect reconstruction.
However, in practice, views are typically taken while the object is rotated around a single fixed axis.

Standard reconstruction methods are based on analytical techniques, such as filtered back projection (FBP) \cite{FBP} for parallel-beam CT and the Feldkamp-Davis-Kress (FDK) algorithm \cite{FDK} for cone-beam CT.
Typically, the large array size of detectors requires the collection of many views so that analytical algorithms can generate high-quality reconstructions.
For example, a detector with dimensions of $2000 \times 2000$ pixels generally needs about 2000 projection measurements at sufficiently high signal-to-noise ratio to achieve a high-quality $2000 \times 2000 \times 2000$ FDK reconstruction \cite{ict_num_view}.

In non-destructive evaluation (NDE) for additive manufacturing (AM), the total scan time can vary from a few hours to a full day, depending on the system's energy and settings, the sample's density and size, and the desired resolution.
However, these long scan times can limit the use of CT in NDE \cite{iCT_time}.
Consequently, AM inspection could greatly benefit from more effective methods of sparse-view acquisition and reconstruction.

To shorten scan times while preserving image quality, researchers have adopted sparse-view reconstruction methods. Techniques such as model-based iterative reconstruction (MBIR) \cite{mbir1,mbir2,mbir3,balke2018separable}, plug-and-play (PnP) algorithms \cite{pnp1,pnp2,pnp3}, and end-to-end deep learning (DL) methods \cite{simurgh,ziabari2023enabling} have shown that high-quality reconstructions can be achieved with far fewer projections compared to the standard methods.

Using sparse-view reconstruction algorithms raises an important question: Given the limited number of projections available, which orientations should be selected to achieve the best reconstruction?
Current workflows typically rely on measurements from a fixed, uniformly sampled set of orientations, without adapting to the unique geometry of the scanned object.
However, empirical studies \cite{bruker2019hartplus,LR2} have shown that object-dependent scanning strategies can outperform uniform sampling in terms of reconstruction quality.
For example, the Bruker High Aspect Ratio Tomography (HART) system \cite{bruker2019hartplus} demonstrates that projections aligned with long, straight edges yield improved resolution in the reconstructed image.
However, it is also crucial that the selected projections remain diverse, rather than all clustering about a few prominent edge features.

While progress has been made in view selection (see Section~\ref{sec:RelatedWork}), determining the optimal views for CT reconstruction remains an open challenge.
In particular, previous research tends to focus on criteria that favor views that are {\em individually most informative}, rather than criteria that favor view sets that are {\em jointly most informative.}
This tends to result in view selection algorithms that do not incorporate the need for view diversity, or alternatively, incorporate diversity through ad-hoc heuristics requiring the manual selection of parameters.

In this paper, we introduce the view covariance loss selection (VCLS) algorithm to jointly select an optimal set of views based on a novel mathematical expression for what we call the view covariance loss (VCL).
The VCL is a measure of the joint information contained in the set of views.
It incorporates into a single loss function both a measure of the information contained in the individual views, as well as a measure of diversity among views.
We derive the VCL from an expression for the normalized mean squared error (NMSE) when the volume is reconstructed using view-basis functions.

The key contributions of this paper are:
\begin{itemize}
    \item Derivation of a novel closed-form parameter-free expression for the VCL loss function that measures the joint information contained in a set of views.
    \item Development of a VCLS algorithm that performs fast joint selection of view angles by approximately minimizing the VCL.
    \item Demonstration of the use and value of VCLS on large practical data volumes that are representative of NDE inspection applications for AM.
\end{itemize}

We apply the VCLS algorithm to the problem of cone-beam CT with a single fixed axis of rotation.
Our approach assumes the availability of a ``reference object'' that can be used for the selection of views.
For example, in AM, many manufactured versions of the same object design must typically be inspected for defects.
In this case, the reference object can come from a scan of one of the many objects being inspected.
However, if every object being manufactured is unique, then the reference object can come from a CAD model, which is typically available, or a quick pre-scan of the object being inspected.
In each case, the VCLS algorithm uses the reference object to select the best view angles to scan during the inspection of the manufactured object(s).

Once the views are collected, we assume the volume is reconstructed using a sparse-view reconstruction algorithm, such as MBIR~\cite{mbirjax-2024}, since it provides higher-quality reconstructions from sparse-view data compared to standard techniques such as FDK.

Experimental results on simulated and measured cone-beam CT data show that the VCLS consistently outperforms both uniform view angle sampling as well as recent {\em ad hoc} strategies that combine view information with view diversity \cite{LREdge3}.
Also, leveraging MBIR reconstructions along with our proposed view angle selection results in fewer artifacts and offers sharper and more accurate edges.

An open source implementation of VCLS is available from \cite{mbirjax-2024} with examples and test data available at \cite{mbirjaxapplications2024}.

\begin{figure}
\centering
\includegraphics[width=0.45\textwidth]{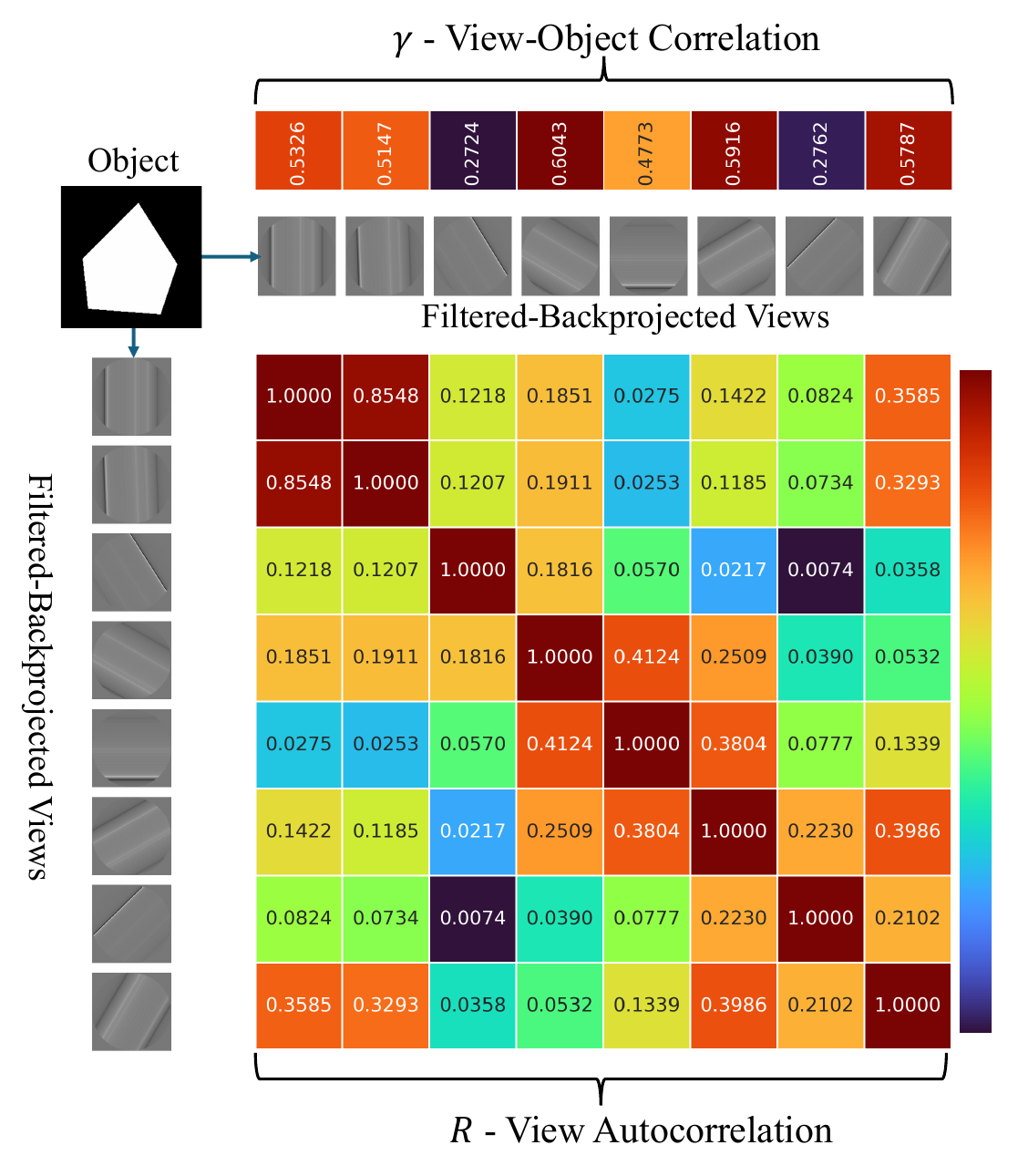}\\
a) Computation of $\gamma$ and $R$\\[15pt]
\includegraphics[width=0.47\textwidth]{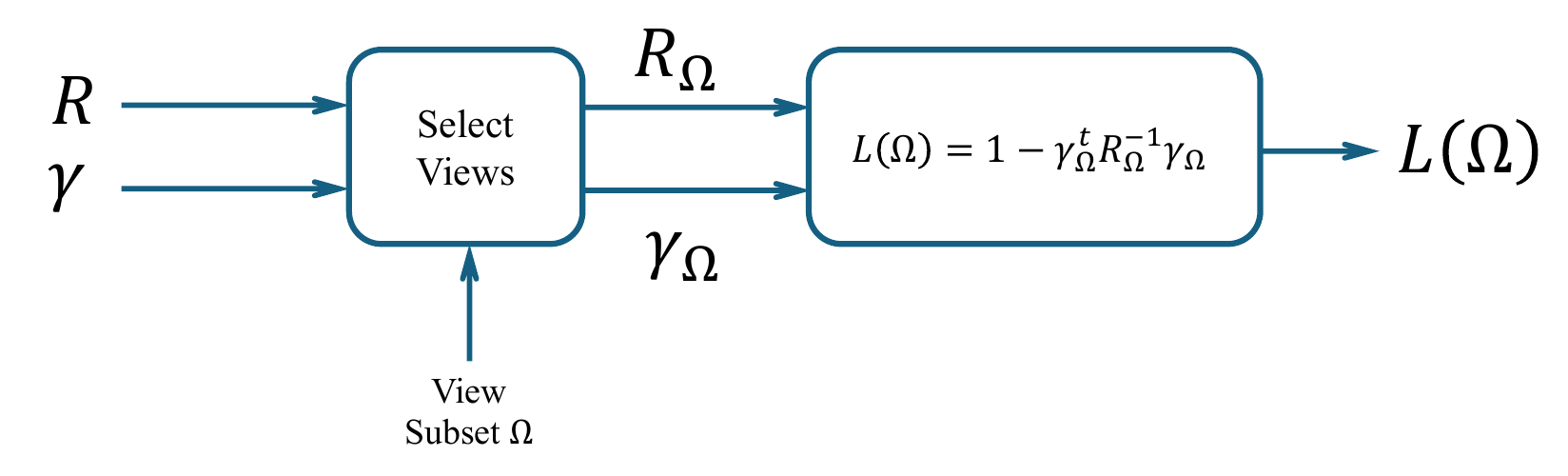}\\
b) Proposed view covariance loss\\[15pt]
\caption{Evaluation of the view covariance loss (VCL) for a toy example of parallel beam tomography with 8 views collected over 180$^\circ$.
a) A typical object to be imaged is forward-projected, filtered, and back-projected to form views in the space domain.
Then $\gamma$ - the cross-correlation between the object and views is computed, and $R$ - the autocorrelation between the views is computed.
b) The $(R,\gamma)$ for the full set of views are subsampled to the selected views, and the resulting $(R_\Omega,\gamma_\Omega)$ are used to evaluate the loss function.
}
\label{fig:fig1}
\end{figure}

\section{Related Work}
\label{sec:RelatedWork}

Over the past 15 years, a variety of approaches have been proposed for object-dependent tomographic view selection.
These approaches seek to select a fixed number of views in a manner that reduces artifacts, improves resolution, and/or reduces noise.

These approaches typically select view angles or other orientation parameters that attempt to maximize the useful information collected with each view.
However, if one uses a completely greedy strategy of view selection, the result will be that too many similar views will be collected.
Consequently, it is essential that views be collected so that they jointly maximize the information.
This inevitably leads to a competing need for diversity in view selection.

For example, Batenburg et al. \cite{LRProj3} considered the problem of view selection for the special case of binary tomography, in which the reconstruction is assumed to take on only two possible densities.
In this case, they assumed that the total information of the projections was inversely proportional to the size of the set of binary solutions.
However, this method does not naturally generalize to the tomographic reconstruction of continuously valued objects.

Several early approaches to view selection are based on the idea that views with higher-frequency content carry more information.
For example, in Haque et al. \cite{LRProj2}, regions of projections with greater spectral richness are allocated a larger number of views.
This non-uniform selection ensures that areas rich in information are sampled more densely.
Similarly, in Presenti et al. \cite{LRProj4 }, the objective is to select views that exhibit the largest intensity differences in the simulated projections generated from a CAD model.

Another related heuristic is that views aligned with the object's edges tend to contain more information, as they capture higher-frequency details and can enhance edge reconstruction quality.
For example, in Matz et al. \cite{LRProj1}, views were selected that aligned with object edges detected using a wavelet transform analysis.
In addition, another metric was employed to measure the angular distance between projections, thereby ensuring diversity in view selection.

In additive manufacturing applications, the edge information of the scanned objects can be directly extracted from the CAD model.
In Yang et al. \cite{LREdge1} and Lin et al. \cite{LREdge2,LREdge3}, algorithms are proposed that construct an edge alignment function to evaluate how well each view aligns with the long edges present in the CAD model, as well as an angle dispersion function to assess the diversity of the views.
By linearly combining these two functions with a weighting parameter $\lambda$, the methods can select views that align with long edges while ensuring sufficient diversity among the chosen views.

Other studies have utilized insights from CT design to improve scanning strategies.
These insights include minimizing the path length through the object to reduce metal artifacts and photon starvation, as well as decreasing the variance of pixel values in a single projection to mitigate beam hardening artifacts.
For example, in Heinzl et al. \cite{LRAtt1}, the authors aim to determine the optimal placement of an object by minimizing the average or maximum path length through the material.
Similarly, in Kano et al. \cite{LRAtt2}, a comparable heuristic is applied, but with enhanced flexibility in controlling cone-beam parameters such as cone angle and tilt angle.
In Ito et al. \cite{LRAtt5}, two metrics are introduced—one for evaluating the path length through the object and another for assessing the variance in a single projection—to help identify an optimal rotation orientation for scanned objects.
However, these approaches overlook the fact that certain orientations may have a long path length through the material yet still contain valuable information, such as long-edge information.

Task-based scanning strategy optimization is also an important research direction.
For example, in \cite{LRDetection1,LRDetection2,LRDetection3,LRDetection4}, methods are proposed to enhance defect detectability within a defined region of interest (ROI).
In \cite{LRDetection4}, this is achieved by optimizing the detectability index within the ROI, which quantifies how well a signal can be distinguished from noise in the frequency domain.
The detectability index is determined by the modulation transfer function and the noise power spectrum of the reconstruction.
These approaches leverage prior knowledge of the ROI to optimize the scanning trajectory, thereby reducing noise and improving image quality in critical areas.
However, these methods are based on prior knowledge of the locations and geometries of defects.

Recent research has explored the use of deep learning for optimal view selection.
For example, in \cite{LRRL1,LRRL2,LRRL3}, deep reinforcement learning (DRL) is employed to sequentially choose the next optimal view, using either projections or reconstructed images as input.
These approaches frame the selection of optimal projections as a trajectory-based optimization problem and solve it using deep reinforcement learning.
The reward function (i.e., the objective optimized during training) is primarily based on reconstruction loss.
However, because of their computational complexity, these approaches have only been demonstrated for low-dimensional images.

In summary, previous research has developed object-dependent scanning strategies that outperform traditional object-independent approaches.
However, aside from DRL-based methods, which do not easily scale to high-dimensional problems, the heuristic-based optimization frameworks used in these studies do not guarantee the selection of an optimal view set that directly minimizes reconstruction loss.
These limitations motivate us to propose a new, computationally efficient loss function that directly guides the selection of views to minimize the reconstruction loss.

\section{The View Covariance Loss (VCL)}
\label{sec:VCL}

We first derive an expression for a loss function that can be used to determine the best view subset of a reference object denoted by $x$.
We assume here that our reference and test objects are similar so that the views selected for the reference object will result in a high-quality scan of the test object.

Depending on the details of the concept of operation, the reference object can be obtained from a scan of object that is similar to the test object, a CAD model of the test object, or a quick low-resolution pre-scan of the test object.
We conclude the section by discussing the interpretation of the loss function and explaining how it can be efficiently computed for CT problems of practical size.

\subsection{VCL Derivation}
\label{sec:VCLDerivation}

Let $\Omega = \{\theta_0, \cdots, \theta_{K-1}\}$ denote a set of scan view parameters that parameterize $K$ tomographic views of an object.
These parameters are typically scan angles in axial tomography, but they can represent anything that parametrizes each view of a scan.

For each view, we will assume that there is a corresponding reconstruction basis function, $T_{\theta_k}\in \mathbb{R}^N$, and that the object, $x$, can be approximately reconstructed as a linear combination of this set of view-basis functions.
Mathematically, we can express this as
\begin{equation}
    \hat{x}_{\Omega, \beta} = \sum_{k=0}^{K-1}{T_{\theta_k}\beta_k}=T_{\Omega}\beta \ ,
    \label{eq:BasisReconstruction}
\end{equation}
where
$$
T_{\Omega}= \left[ T_{\theta_0}, \ldots , T_{\theta_{K-1}} \right] \ ,
$$
is an $N\times K$ matrix of view-basis functions and $\beta \in \mathbb{R}^K$ is a column vector of weights.
Here, $N$ is the number of voxels in the reconstruction and $K$ is the number of views used to perform the reconstruction.

The selection of view-basis functions depends on the application.
For example, in our CT reconstruction problem, if an analytical reconstruction algorithm such as FBP or FDK is used, then the reconstructed image is computed as a linear combination of filtered back-projections \cite{hsieh2021ct}.
In this case, each view-basis function is chosen to be the filtered back projection (FBP) from a single view at angle $\theta_k$
\begin{equation} \label{eq:reconbasis}
    T_{\theta_{k}} = \alpha_{\theta_{k}} \text{FBP}(A_{\theta_{k}}x) = \alpha_{\theta_{k}} A_{\theta_{k}}^t H A_{\theta_{k}}x \ ,
\end{equation}
where $A_{\theta_{k}}x$ represents the simulated forward projection of the reference object, $x$, at angle $\theta_{k}$, $H$ represents the required ramp filter that also depends on the choice of geometry \cite{FBP,FDK}, and $\alpha_{\theta_{k}}$ is any fixed scaling constant.

Based on this setup, our goal is to obtain the optimal reconstruction by minimizing the normalized mean squared error (NMSE) loss between $x$ and $\hat{x}$ through the selection of the optimal subset given by
\begin{equation}
\Omega^\ast = \argmin_{\Omega}{
    \left\{ \min_{\beta} \dfrac{\|x-\hat{x}_{\Omega, \beta}\|^2}{\|x\|^2} \right\} } \ ,
\end{equation}
where the reconstruction, $\hat{x}_{\Omega, \beta}$, depends on the selection of the set of view angles, $\Omega$, and weights, $\beta$.

From this, we can see that our goal is to find the set of view angles, $\Omega^\ast$, that minimizes the reconstruction distortion as shown below
\begin{equation}
\Omega^\ast = \argmin_{\Omega}{ D( \Omega ) } \ ,
\label{eq:OptimalViewSubset}
\end{equation}
where $D( \Omega )$ is the NMSE in the reconstruction given by
\begin{equation}
\label{eq:NRMSEDistortion}
D( \Omega ) = \min_{\beta } \dfrac{\|x- \hat{x}_{\Omega, \beta} \|^2}{\|x\|^2} \ .
\end{equation}

We can simplify the form of $D( \Omega )$ by calculating the optimal vector $\beta_{\Omega}^\ast$ for a specific choice of $\Omega$.
This is given by the solution to the following least-squares estimation problem.
\begin{equation*}
    \beta_{\Omega}^\ast = (T_{\Omega}^tT_{\Omega})^{-1}T_{\Omega}^tx
\end{equation*}
Substituting $\beta_{\Omega}^\ast$ into~\eqref{eq:BasisReconstruction} and~\eqref{eq:NRMSEDistortion} then results in
\begin{equation*}
D( \Omega ) = \dfrac{\|x-P_{\Omega }x\|^2}{\|x\|^2}
\end{equation*}
where $P_{\Omega }=T_{\Omega}(T_{\Omega}^tT_{\Omega})^{-1}T_{\Omega}^t$.
Note that $P_{\Omega }$ performs an orthogonal projection onto the column space of $T$.
Consequently, the residual $x-P_{\Omega } x$ is orthogonal to the column space of $T$.
Therefore, the norm squared $\|x-P_{\Omega }x\|^2$ can be computed using Pythagorean theorem
\begin{equation}
    \|x-P_{\Omega }x\|^2 = \|x\|^2-\|P_{\Omega }x\|^2
    \label{eq:PythagoreanTheorem}
\end{equation}
where $\|P_{\Omega } x\|^2$ can be simplified as
\begin{align}
\nonumber
    \|P_{\Omega } x\|^2 &= x^t P_{\Omega }^t P_{\Omega } x \\
    &= x^t T_{\Omega} (T_{\Omega}^t T_{\Omega})^{-1} T_{\Omega}^t x \ ,
    \label{eq:ProjectionNormSquared}
\end{align}
and we use the fact that $P_{\Omega }^t P_{\Omega }=P_{\Omega }$ because $P_{\Omega }$ is an idempotent matrix.
We can next use~\eqref{eq:PythagoreanTheorem} and~\eqref{eq:ProjectionNormSquared} to derive the following simplified expression for the NMSE.
\begin{align*}
D( \Omega )
    & = \dfrac{\|x - P_{\Omega } x\|^2}{\|x\|^2} \\
    &= \dfrac{\|x\|^2 - \|P_{\Omega } x\|^2}{\|x\|^2} \\
    &= 1 - \dfrac{x^t T_{\Omega} (T_{\Omega}^t T_{\Omega})^{-1} T_{\Omega}^t x}{\|x\|^2}
\end{align*}

From this result, we can now write the NMSE reconstruction error as
\begin{equation}
D( \Omega ) = L( R_{\Omega}, \gamma_{\Omega} )
\label{eq:ViewCovarianceLoss}
\end{equation}
where $L$ is the view covariance loss (VCL) function defined by
\begin{align}
\label{eq:ViewSelectionLoss}
L( R, \gamma ) &= 1 - \gamma^t R^{-1} \gamma  \ .
\end{align}
and its arguments are given by
\begin{align}
\label{eq:CorrelationMatrix}
R_{\Omega} &= T_{\Omega}^t T_{\Omega} \\
\gamma_{\Omega} &= \dfrac{T_{\Omega}^t x}{\|x\|} \ ,
\end{align}
where $R_{\Omega}$ is the autocorrelation between the reconstruction bases, and $\gamma$ is the correlation between the view bases and the reference object.

\subsection{VCL Computation}
\label{sec:VCLDerivationComputation}

Figure~\ref{fig:fig1} illustrates how the VCL is computed for a simple example of 8-view parallel-beam tomography.
First, the filtered back-projection is computed for each view of the object.
Then the matrix $R$ is formed by the autocorrelation of the back-projected views, and the vector $\gamma$ is formed by the cross-correlation between the views and the object.
In the figure, we assume that $\alpha_{\theta} = 1/\| T_\theta \|$ in \eqref{eq:reconbasis} so that the reconstruction bases are normalized.
Consequently, the entries of both $R$ and $\gamma$ are normalized correlations so they take values in the range $[-1,1]$, and $R$ has 1's on its diagonal.

Here, $R\in \mathbb{R}^{K_{max} \times K_{max}}$  and $\gamma \in \mathbb{R}^{K_{max}}$ where $K_{max}$ corresponds to the largest set of views to be considered.
Even for large practical problems, typically $K_{max} \leq 2000$.
Consequently, $(R,\gamma )$ can be easily stored and used in calculations.

However, naive calculation of $(R,\gamma )$ from the back-projected views can be very computationally expensive.
This is because for large practical problems, each back projection can be very large; so computing and storing these back projections may not be practical.
For example, for a reference object with an array size of \(1024 \times 1024 \times 1024\) and 1000 candidate views, computing \(R\) requires evaluating approximately \(5 \times 10^5\) inner products between two matrices, each with dimensions \(1024 \times 1024 \times 1024\).

Fortunately, the computation and memory requirements for computing $(R,\gamma)$ can be dramatically reduced by subsampling the reconstruction bases.
Computing a single entry of \(R\) requires taking the inner-product of two very large reconstruction bases.
However, sufficient accuracy can be achieved by computing the inner product using only a small randomly selected subset of voxels.
Let $S\in \mathbb{R}^{N_S \times N}$ be a random sub-sampling matrix where $N_S<<N$ is the number of sample voxels.
In our experiments, we will choose $N_S= \lfloor r_1 N \rfloor$ where $r_1$ is a sampling rate parameter typically chosen to be very small.
So then to compute the VCL using \eqref{eq:ViewSelectionLoss} and
\begin{align}
R &= T^t S^t S T \label{eq:R_computation} \\
\gamma &= \frac{ T^t S^t S x }{ \| Sx \| } \ ,
\label{eq:gamma_computation}
\end{align}
where $(R,\gamma)$ are computed using the full set of possible views to be considered.

Figure~\ref{fig:fig1}(b) illustrates how $(R_\Omega,\gamma_\Omega)$ are computed by selecting the rows and columns of $(R,\gamma)$ corresponding to the chosen subset, $\Omega$, of the complete set of $K_{max}$ views.
The value of VCL is then computed from $(R_\Omega,\gamma_\Omega)$.
Notice that the computation of VCL requires a matrix inversion so it is ${\cal O} ( K^3 )$, but since $K < K_{max}$, this computation is not expensive.

\subsection{VCL Interpretation}
\label{sec:VCLIntuition}

In order to better understand the VCL, consider a measurement model with the form
$$
y = T^t (\tilde{x} + \epsilon ) \ ,
$$
where $y$ is the vector of observations, $T$ is the matrix of bases, $\tilde{x}$ is the unknown object, and $\epsilon \sim {\cal N}(0,I)$ is white Gaussian noise with unit variance.
Furthermore, define a linear function $\tilde{x} = c \, \eta$ where $\eta \in \mathbb{R}$ and $c\in \mathbb{R}^N$.

Then in the Supplementary Material, we show that the Fisher information for the estimation of $\tilde{x}$ given $y$ is given by
$$
{\cal I}_{\tilde{x}} = T R^{-1} T^t \ ,
$$
and the Fisher information for the estimation of $\eta$ is given $y$ is given by
$$
{\cal I}_{\eta} = c^t T R^{-1} T^t c \ .
$$
Then if we choose $c=x/\|x\|$ where $x$ is the reference object, we get the following expression for the Fisher information
$$
{\cal I}_{VCL} = \gamma^t R^{-1} \gamma = 1 - VCL \ .
$$
So, from this we can see that minimizing the VCL has the interpretation of maximizing the Fisher information for the estimation of the scalar quantity $\eta$ when $\tilde{x} = x \eta$ where $x$ is the reference object and $\tilde{x}$ is the unknown reconstruction.

This Fisher information can also be interpreted as maximizing the quadratic form given by
\begin{equation}
{\cal I}_{VCL} = \gamma^t B \gamma \ ,
\label{eq:QuadraticForm}
\end{equation}
where $B = R^{-1}$ is a precision matrix.

So this raises the question of how to interpret maximization of the view information, ${\cal I}_{VCL}$?
First, maximizing the components of $\gamma_{\Omega}$ will increase the view information.
This is intuitive, since a view that is correlated with $x$ should collect more information about $x$.

The matrix $B = R^{-1}$ can be interpreted as a precision matrix that encodes the inverse covariance of the view-basis functions.
Without loss of generality, we can again assume that the reconstruction bases are normalized so that, $\| T_{\theta_k}\|=1$, in which case $\mbox{diag}\{ R_{\Omega} \}=I$.
Then the off-diagonal elements of $B$ reflect the degree to which different views can be predicted from each other.
Large negative off-diagonal values indicate strong mutual predictability; so $B$ acts to penalize redundancy among views.

Consequently, maximizing ${\cal I}_{VCL}$ favors the selection of views that are both informative and diverse, which is precisely what view selection should do.
Importantly, VCL incorporates both information and diversity in a natural way without requiring separate {\em ad hoc} weighted functions for these two purposes.

\begin{algorithm}
\caption{View Covariance Loss Selection Algorithm}
\begin{algorithmic}
\STATE \textbf{Function}: $\text{VCLS}(\Omega,x;K,r_1,r_2)$
\STATE \textbf{Output}: $\Omega^\ast$
\STATE // Step 1: Compute the reconstruction bases
\STATE $S \leftarrow \text{RandSubsampleMatrix}(x,r_1)$
\STATE for $i$ in $1$ to $|\Omega|$ do:
\STATE \hspace{0.5cm} $T_{\theta_i} \leftarrow A_{\theta_i}^t H A_{\theta_i}x$
\STATE \hspace{0.5cm} $T_{\theta_i} \leftarrow S T_{\theta_i} / \|S T_{\theta_i}\|$
\STATE $x \leftarrow S x$
\STATE // Step 2: Compute $(R,\gamma)$
\STATE for $i$ in $1$ to $|\Omega|$ do:
\STATE \hspace{0.5cm} $\gamma[i] \leftarrow T_{\theta_i}^t x / \|x\|$
\STATE \hspace{0.5cm} for $j$ in $1$ to $|\Omega|$ do:
\STATE \hspace{1.0cm} $R[i][j] \leftarrow T_{\theta_i}^t T_{\theta_j}$
\STATE // Step 3: Optimized View Selection
\STATE $\Omega^\ast \leftarrow \text{ViewSubsetSelection} (\Omega,R,\gamma;K,r_2)$
\STATE \textbf{return} $\Omega^\ast$
\end{algorithmic}
\label{alg1}
\end{algorithm}

\begin{algorithm}
\caption{View Subset Selection}
\begin{algorithmic}
\STATE \textbf{Function}: $\text{ViewSubsetSelection}(\Omega,R,\gamma;K,r_2)$
\STATE \textbf{Output}: $\Omega^\ast$
\STATE // Initialize with uniformly sampled view angles
\STATE $\Omega^\ast \leftarrow \text{UniformSampling}(K,\Omega)$
\STATE // Compute the initial loss value using Algorithm \ref{alg3}
\STATE $L^\ast \leftarrow \text{VCL}(R,\gamma,\Omega^\ast)$
\STATE // Test random view swaps
\STATE Repeat until no change:
\STATE \hspace{0.5cm} for $\theta_i \in \Omega^\ast$ do:
\STATE \hspace{1.0cm} $\Omega_S \leftarrow \text{RandomSampling}(r_2,\Omega - \Omega^\ast)$
\STATE \hspace{1.0cm} for $\theta_j \in \Omega_S$ do:
\STATE \hspace{1.5cm} $\Omega^\prime \leftarrow \text{SwapAngle}(\Omega^\ast,\theta_i,\theta_j)$
\STATE \hspace{1.5cm} $L^\prime \leftarrow \text{VCL}(R,\gamma,\Omega^\prime)$
\STATE \hspace{1.5cm} if $L^\prime < L^\ast$ do
\STATE \hspace{2.0cm} $\Omega^\ast \leftarrow \Omega^\prime$
\STATE \hspace{2.0cm} $L^\ast \leftarrow L^\prime$
\STATE \textbf{return} $\Omega^\ast$
\end{algorithmic}
\label{alg2}
\end{algorithm}

\begin{algorithm}
\caption{Compute View Covariance Loss}
\begin{algorithmic}
\STATE \textbf{Function}: $\text{VCL}(R, \gamma, \Omega)$
\STATE \textbf{Output}: $L$
\STATE // Subsample
\STATE $R_{\Omega} \leftarrow \text{Subsampling}(R,\Omega)$
\STATE $\gamma_{\Omega} \leftarrow \text{Subsampling}(\gamma,\Omega)$
\STATE // Compute Loss
\STATE $L \leftarrow 1 - \gamma_{\Omega}^t R_{\Omega}^{-1} \gamma_{\Omega}$
\STATE \textbf{return} $L$
\end{algorithmic}
\label{alg3}
\end{algorithm}

\begin{figure}
\centering
\includegraphics[width=0.4\textwidth]{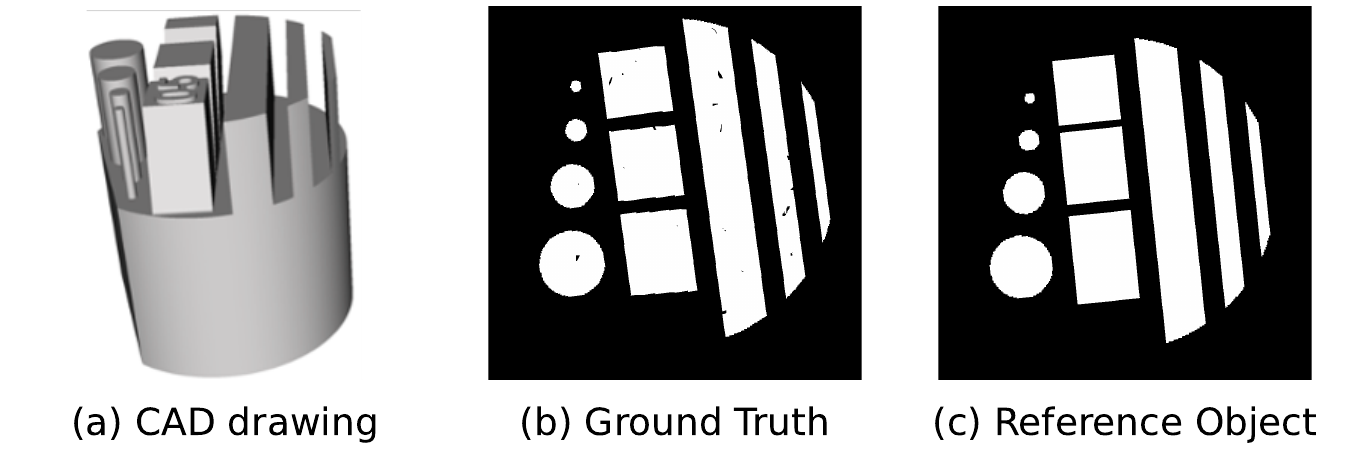}
\caption{
    (a) The CAD drawing used in the simulated data experiment; (b) axial slice of ground truth; and (c) axial slice of reference object used in view selection.
    Note that reference object is not identical to ground truth.
}
\label{fig:simulated_train_object}
\end{figure}

\begin{table}
\caption{Parameters for Simulated CT Experiment}
\vspace*{-5pt}
\centering
\renewcommand{\arraystretch}{1.2}
\begin{tabular}{|c|c|}
\hline
\textbf{Specification} & \textbf{Value} \tabularnewline
\hline
Geometry & cone-beam \tabularnewline
\hline
Detector Dimensions (pixels)  & $1000 \times 1000$ \tabularnewline
\hline
Pixel Pitch in Detector (mm)  & $0.127$ \tabularnewline
\hline
Source-to-Detector Distance (mm)  & $808.793$ \tabularnewline
\hline
Source-to-Center of Rotation Distance (mm)  & $110.044$ \tabularnewline
\hline
Maximum Energy of X-ray Source (keV)  & $200$ \tabularnewline
\hline
Ground Truth Material  & Iron \tabularnewline
\hline
MBIRJAX Params - $(\text{sharpness},\text{snr\_db})$ & $(1.0,35.0)$ \tabularnewline \hline
Recon Array Size (slice first) & $1000 \times 1000 \times 1000 $ \tabularnewline \hline
Voxel Pitch in Object (mm)  & $0.017$ \tabularnewline \hline
VCLS Params (default)  & $r_1 = 0.001$, $r_2 = 0.1$ \tabularnewline \hline
\end{tabular}
\label{table:parameters}
\end{table}

\begin{table*}
\caption{VCLS Comparison for Different $(r_1, r_2)$ with $K=100$ and $K_{max}=720$}
\vspace*{-5pt}
\centering
\renewcommand{\arraystretch}{1.2}
\begin{tabular}{|c|c|c|c|c|}
\hline
 & \textbf{$r_1=1.0,r_2=1.0$}
 & \textbf{$r_1=0.01,r_2=0.5$}
 & \textbf{$r_1=0.001,r_2=0.1$}
 & \textbf{$r_1=0.0001,r_2=0.01$} \\
\hline
Step1: compute reconstruction bases  &  11.60 minutes &  8.13 minutes  & 5.07 minutes &  4.64 minutes  \\
\hline
Step2: Compute $(R,\gamma)$           & 456.01 minutes & 7.24 minutes   & 1.51 minutes &  0.26 minutes \\
\hline
Step3: Compute View Subset Selection     & 2.93 minutes &  2.21 minutes   & 0.39 minutes &  0.04 minutes \\
\hline \hline
\textbf{Total Time} & 470.54 minutes & 17.58 minutes & 6.97 minutes & 4.94 minutes \\
\hline
\textbf{Reconstruction Error (NRMSE)} & 0.0780 & 0.0792 & 0.0794 & 0.0801 \\
\hline
\end{tabular}
\label{table:time}
\end{table*}

\begin{figure*}
    \centering
    \begin{subfigure}[b]{0.65\textwidth}
     \centering\hspace*{-15pt}
     \includegraphics[width=1.05\textwidth]{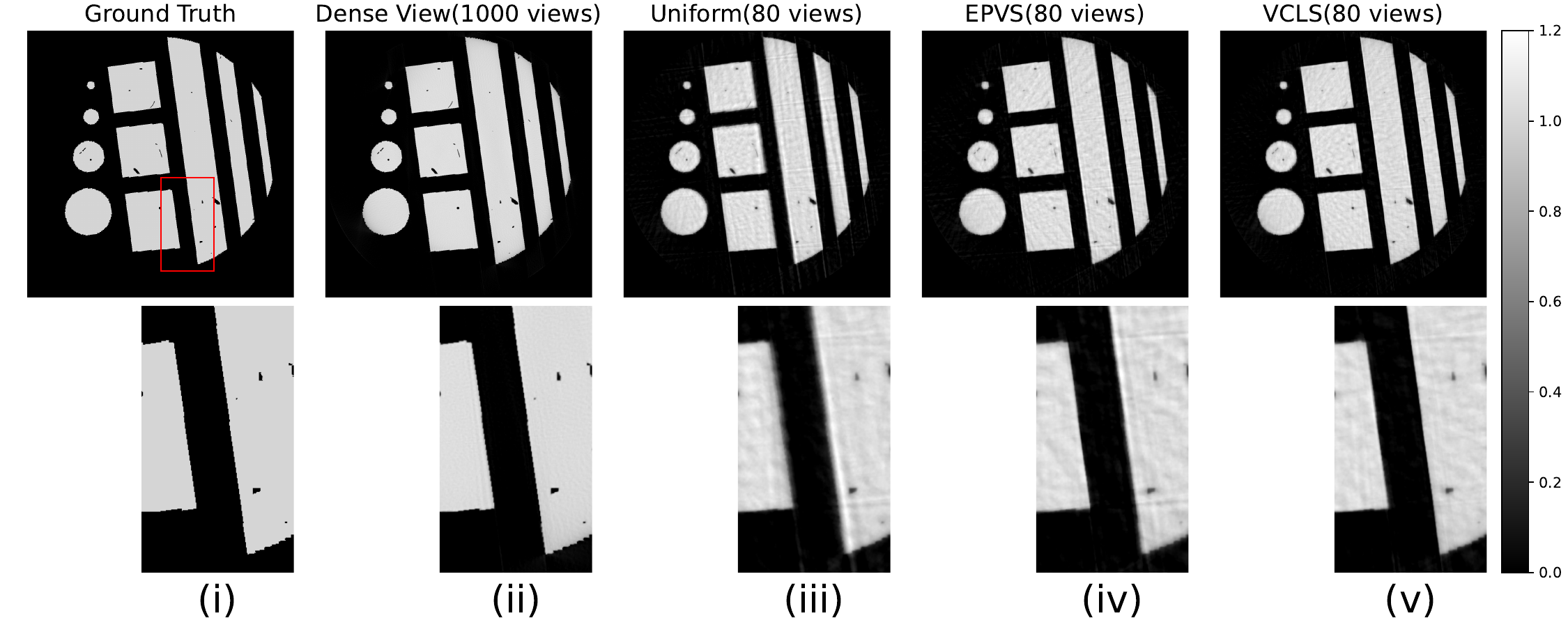}
     \caption{Reconstruction}
     \label{fig:recon(simulated)}
     \end{subfigure}
     \begin{subfigure}[b]{0.3\textwidth}
         \centering
         \includegraphics[width=1.2\textwidth]{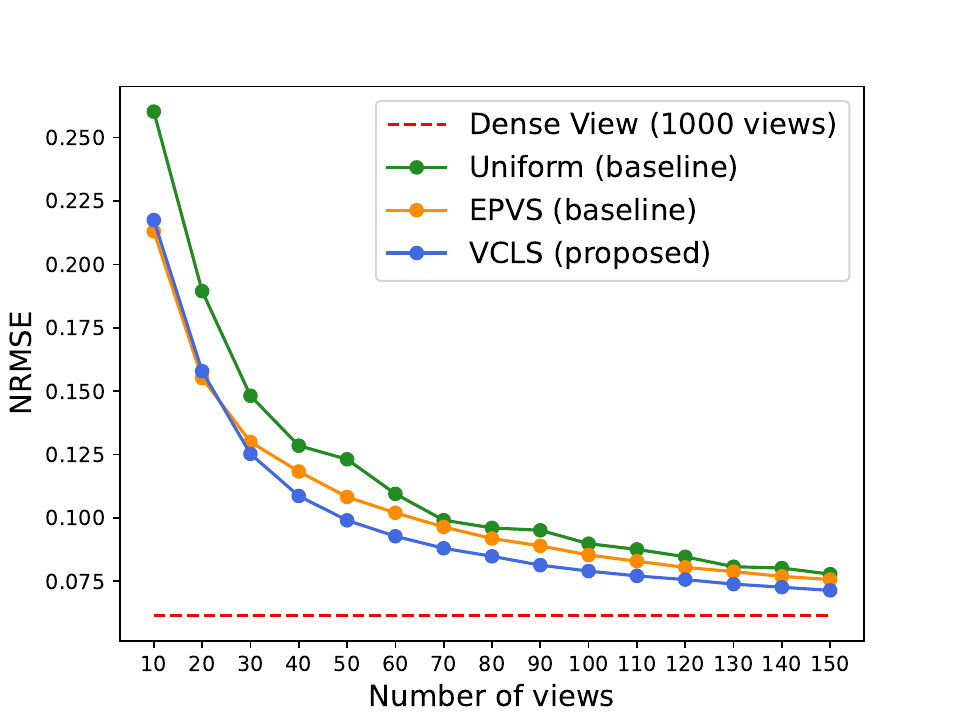}
         \caption{Reconstruction loss}
         \label{fig:nrmse}
     \end{subfigure}
    \caption{
    Comparison of view selection methods using simulated cone-beam axial CT data.
    (a) Axial MBIR reconstructions with the top row being the full slice and the bottom row being zoomed-in portion in red rectangle.
    (i) Ground truth; (ii) 1000 uniformly sampled views; (iii) 80 uniformly sampled views; (iv) 80 EPVS sampled views; and (v) 80 VCLS sampled views.
    (b) Plot of NRMSE for VCLS, EPVS, and uniform view sampling.
    Notice that VCLS produces consistently low NRMSE with sharper edges and more refined texture detail than either uniform sampling or EPVS.
    }
\end{figure*}

\begin{table}
\caption{Parameters for fuel nozzle CT experiment}
\vspace*{-5pt}
\centering
\renewcommand{\arraystretch}{1.2}
\begin{tabular}{|c|c|}
\hline
\textbf{Specification} & \textbf{Value} \tabularnewline
\hline
Instrument  & ZEISS Metrotom 800 \tabularnewline \hline
Geometry  & cone-beam \tabularnewline \hline
Detector Dimensions (pixels)  & $1024 \times 1024$ \tabularnewline
\hline
Pixel Pitch in Detector (mm)  & $0.2$ \tabularnewline
\hline
Source-to-Detector Distance (mm)  & $808.508$ \tabularnewline
\hline
Source-to-Center of Rotation Distance (mm)  & $243.307$ \tabularnewline
\hline
Maximum Energy of X-ray Source (keV)  & $180$ \tabularnewline
\hline
Scanned Object Material  & 316L stainless steel \tabularnewline
\hline
MBIRJAX Params - $(\text{sharpness},\text{snr\_db})$ & $(1.0,35.0)$ \tabularnewline \hline
Recon Array Size (slice first)  & $893 \times 1008 \times 1008$ \tabularnewline \hline
Voxel Pitch in Object (mm)  & $0.06$ \tabularnewline \hline
VCLS Params (default)  & $r_1 = 0.001$, $r_2 = 0.1$ \tabularnewline \hline
\end{tabular}
\label{table:measured_parameters}
\end{table}

\begin{figure}
\centering
\includegraphics[width=0.3\textwidth]{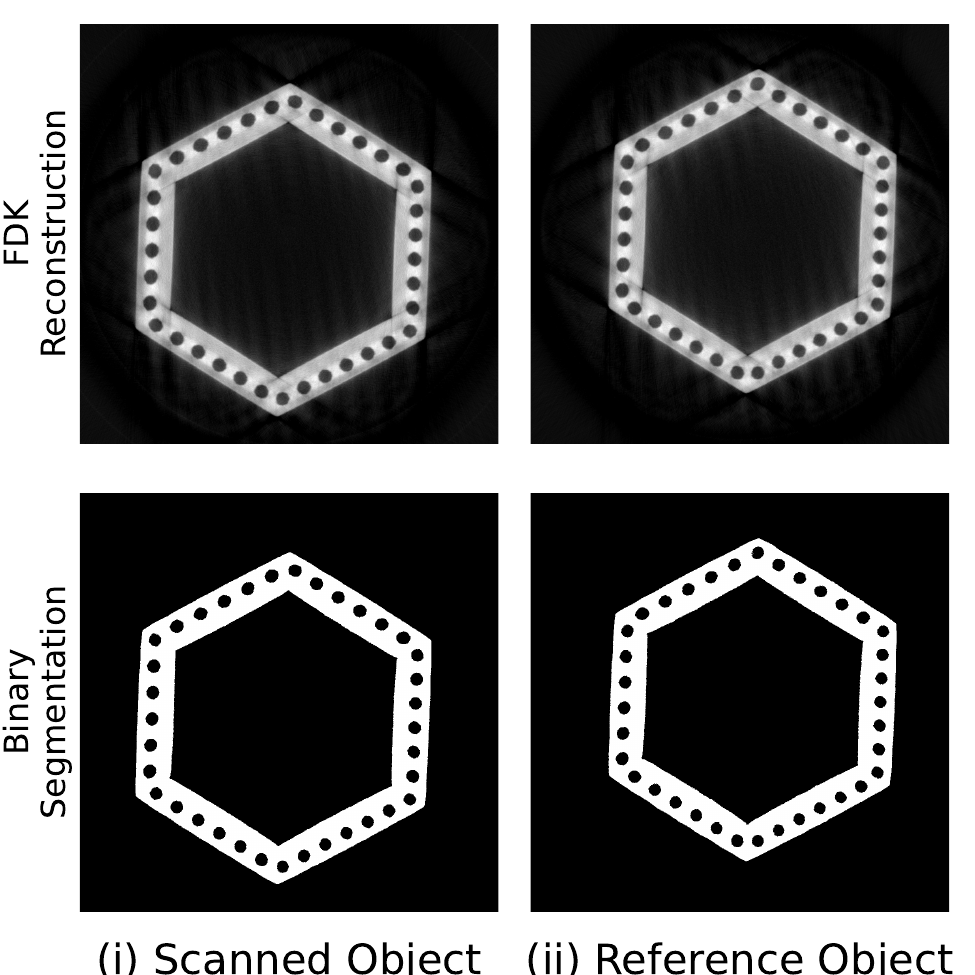}
\caption{
Hexagonal fuel nozzle axial slice for (i) the object scanned in experiments; (ii) the reference object used for view selection.
Top row: FDK reconstructions from dense view sampling.
Bottom row: Segmentation of FDK reconstructions.
The segmentation of the reference object was used as $x$ in VCL evaluation.
Notice that the scanned object and the reference object are not exactly the same with a slight positional shift and some differences in details.
}
\label{fig:measured_training_object}
\end{figure}

\begin{figure*}
\centering\hspace*{-5pt}
\includegraphics[height=1.75in]{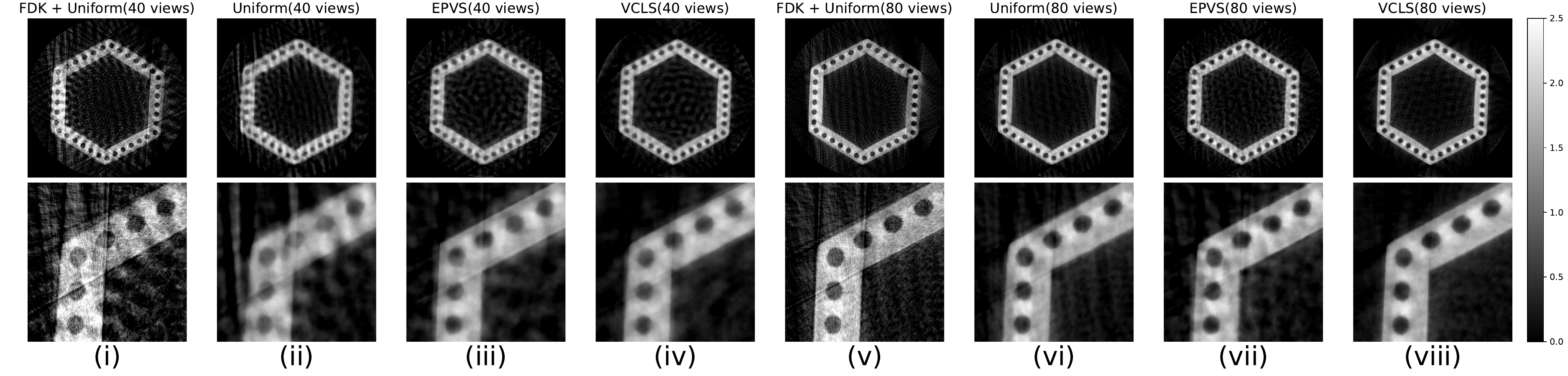}
\caption{Comparison of view selection methods using hexagonal fuel nozzle data.
Top row is the full axial slice of reconstruction, and bottom row is a zoomed-in portion of the region outlined in red in Figure~\ref{fig:LineProfile}(b).
(i) $40$ uniformly-sampled views + FDK (baseline method);
(ii) $40$ uniformly-sampled views + MBIR (baseline method);
(iii) $40$ EPVS sampled views + MBIR (baseline method);
(iv) $40$ VCLS sampled views + MBIR (proposed method).
(v) $80$ uniformly-sampled views + FDK (baseline method);
(vi) $80$ uniformly-sampled views + MBIR (baseline method);
(vii) $80$ EPVS sampled views + MBIR (baseline method);
(viii) $80$ VCLS sampled views + MBIR (proposed method).
Notice that VCLS produces reconstructions with sharper edges and fewer artifacts.
}
\label{fig:MeasuredRecon}
\end{figure*}

\begin{figure}
\centering
\begin{minipage}{\columnwidth}
    \centering \hspace*{-10pt}
    \includegraphics[width=\textwidth]{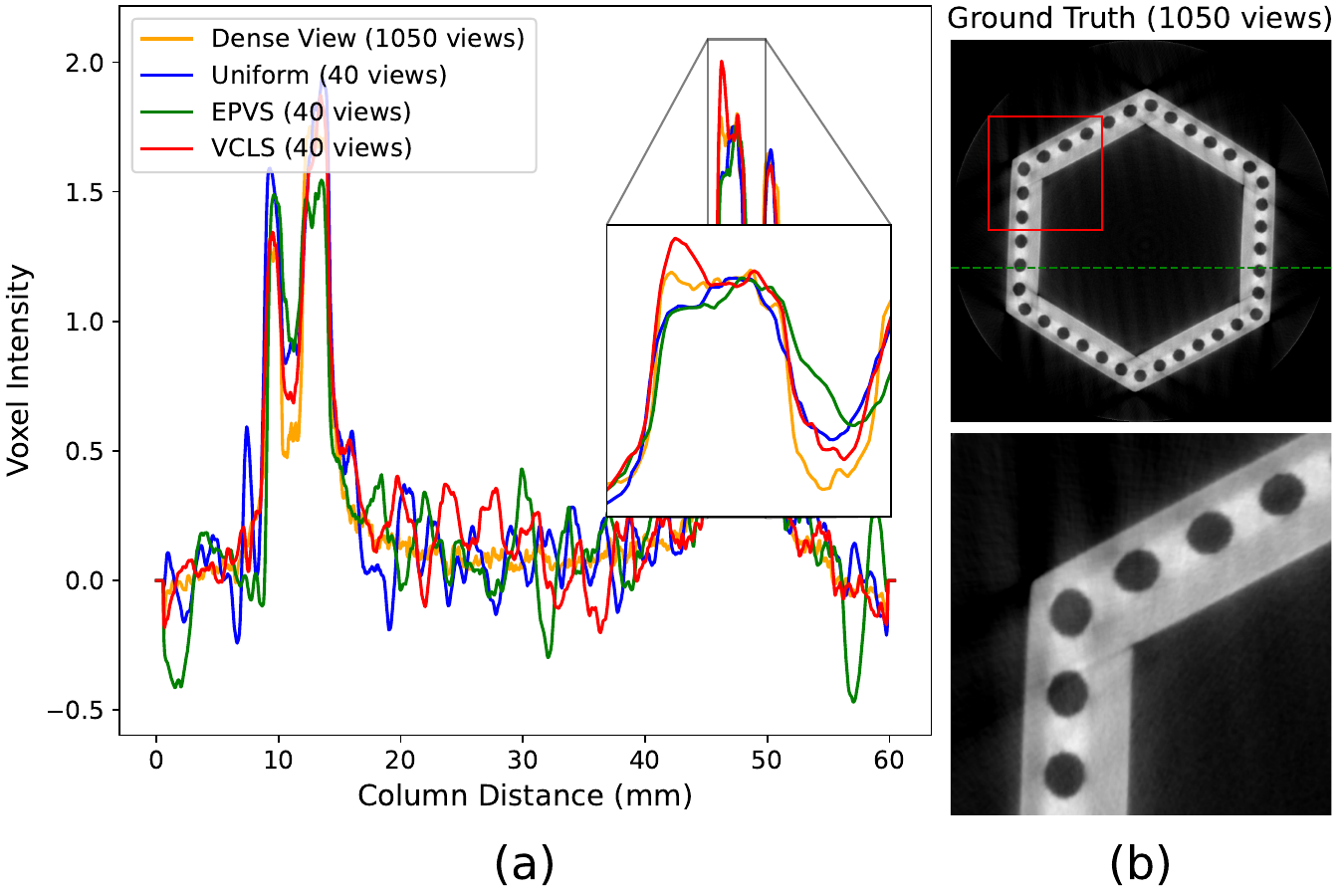}
    \caption{Hexagonal fuel nozzle (a) intensity profiles; (b) ground truth from MBIR reconstruction using 1050 uniformly sampled views.
    Red box indicates zoomed-in region and green line indicates line used for intensity profile.}
    \label{fig:LineProfile}
\end{minipage}
\end{figure}

\begin{table}
\caption{FWHM widths from fuel nozzle data.}
\vspace*{-5pt}
\centering
        \begin{tabular}{|c|c|c|c|c|c|c|}
        \hline &
        \multicolumn{2}{c|}{\textbf{Uniform}} &
        \multicolumn{2}{c|}{\textbf{EPVS}} &
        \multicolumn{2}{c|}{\textbf{VCLS}} \\ \hline
        \makecell{num.\\ views} & \makecell{width\\(mm)} & \makecell{error\\(mm)} & \makecell{width\\(mm)} & \makecell{error\\(mm)} & \makecell{width\\(mm)} & \makecell{error\\(mm)} \\ \hline
        1050  & 2.34 & 0 & \diagbox{}{} & \diagbox{}{} & \diagbox{}{} & \diagbox{}{}  \\ \hline
        40  & 2.70 & 0.36 & 3.18 & 0.84 & 2.46 & {\bf 0.12} \\ \hline
        80  & 2.46 & 0.12 & 1.92 & -0.42 & 2.34 & {\bf 0.00}  \\ \hline
        \end{tabular}
\label{table:FWHM}
\end{table}

\begin{table}
\caption{Hole diameters from fuel nozzle data.}
\vspace*{-5pt}
\centering
    \begin{tabular}{|c|c|c|c|c|c|c|}
    \hline &
    \multicolumn{2}{c|}{\textbf{Uniform}} &
    \multicolumn{2}{c|}{\textbf{EPVS}} &
    \multicolumn{2}{c|}{\textbf{VCLS}} \\ \hline
    \makecell{num.\\ views} & \makecell{dia.\\(mm)} & \makecell{error\\(mm)} & \makecell{dia.\\(mm)} & \makecell{error\\(mm)} & \makecell{dia.\\(mm)} & \makecell{error\\(mm)} \\ \hline
    1050  & \makecell{1.89\\ $\pm$ 0.05} & 0 & \diagbox{}{} & \diagbox{}{} & \diagbox{}{} & \diagbox{}{}  \\ \hline
    40  & \makecell{1.08\\ $\pm$ 0.47} & 0.81 & \makecell{1.34\\ $\pm$ 0.30} & 0.55 & \makecell{1.59\\ $\pm$ 0.11} & {\bf 0.30} \\ \hline
    80 & \makecell{1.76\\ $\pm$ 0.08} & 0.13 & \makecell{1.77\\ $\pm$ 0.06} & 0.12 & \makecell{1.79\\ $\pm$ 0.06} & {\bf 0.10}  \\ \hline
    \end{tabular}
\label{table:DiameterTable}
\end{table}

\begin{figure}
  \centering
  \begin{subfigure}[b]{0.22\textwidth}
    \centering
    \includegraphics[width=1\textwidth]{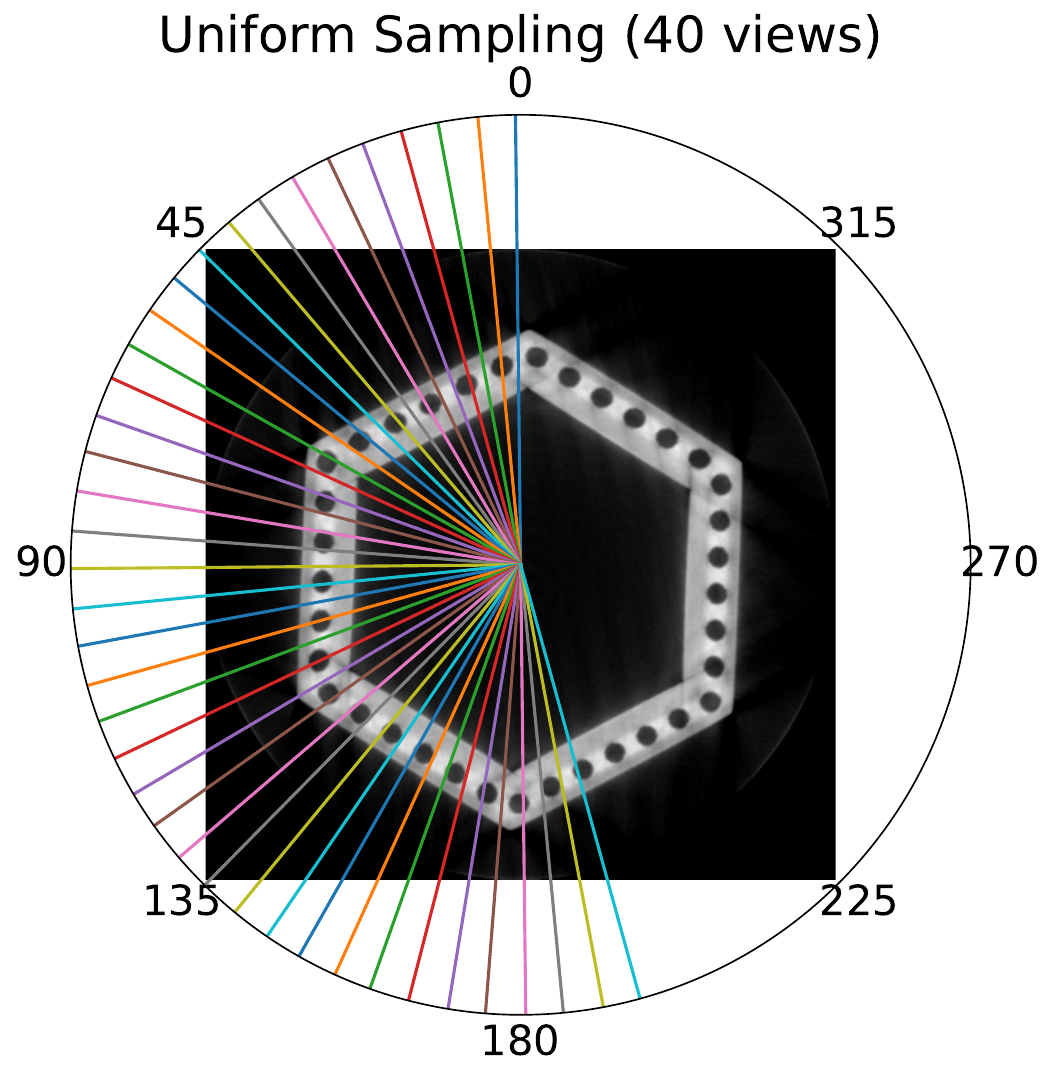}
    \caption{uniform}
  \end{subfigure}
  \hfill
  \begin{subfigure}[b]{0.22\textwidth}
    \centering
    \includegraphics[width=1\textwidth]{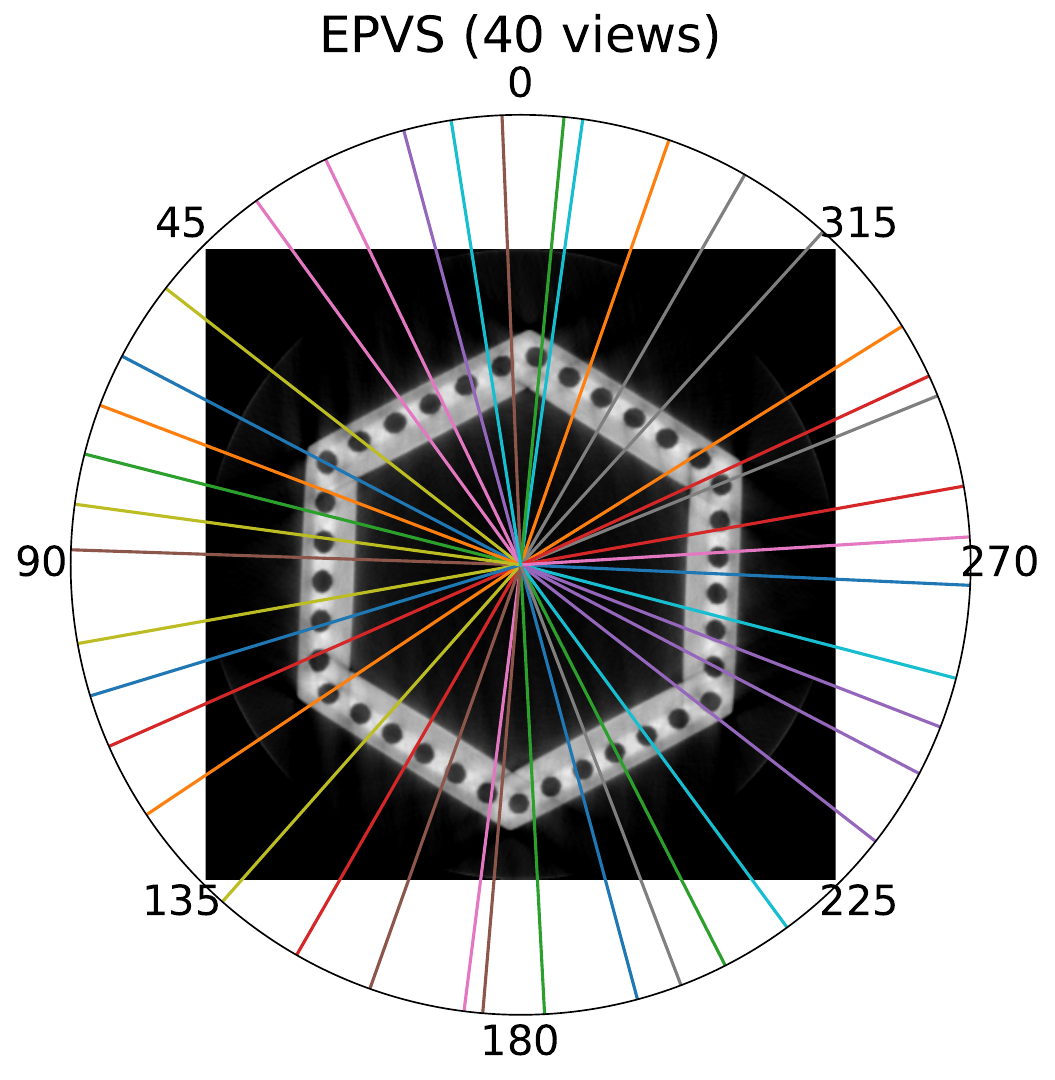}
    \caption{EPVS}
  \end{subfigure}
  \hfill
  \begin{subfigure}[b]{0.22\textwidth}
    \centering
    \includegraphics[width=1\textwidth]{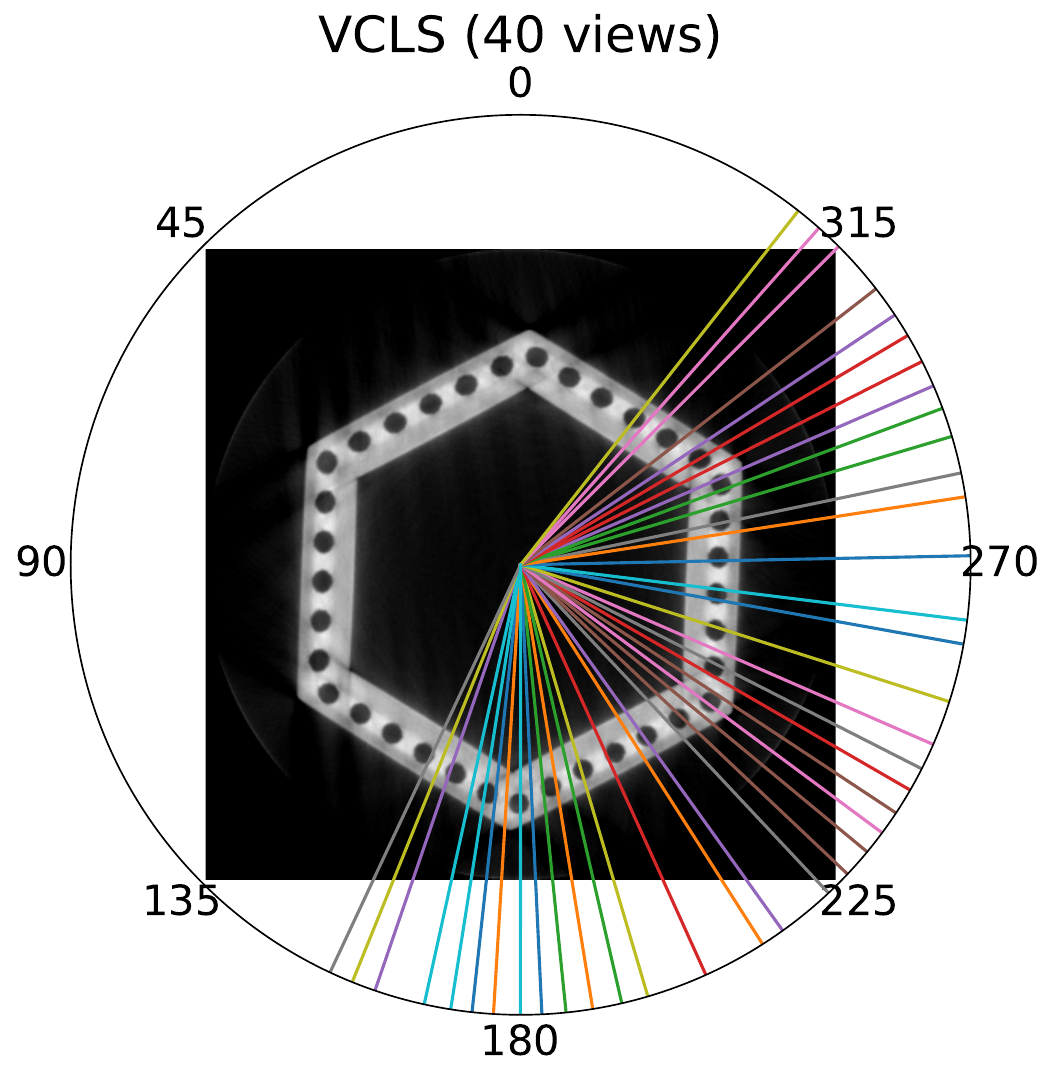}
    \caption{VCLS}
  \end{subfigure}
  \caption{Illustration of the 40 selected views using radial lines for hexagonal fuel nozzle data using (a) uniform sampling (short scan), (b) EPVS, and (c) VCLS.
  Notice both EPVS and VCLS select views with non-uniform spacing.
  However, VCLS automatically selects views with a narrower angular range (less than 180°).}
  \label{fig:angle_selection}
\end{figure}

\section{View Covariance Loss Selection (VCLS) }
\label{sec:implementation}

In this section, we describe an implementation for the View Covariance Loss Selection (VCLS) algorithm which approximately minimizes the VCL in order to select the best view subset.
The general strategy of the VCLS algorithm is to first compute $(R,\gamma)$ for the full set of $K_{max}$ possible views.
The algorithm then searches through possible view subsets of size $K$ to find the subset $\Omega^\ast$ that approximately minimizes $L(R_{\Omega^\ast},\gamma_{\Omega^\ast})$.
Since this requires a search through ${K_{max} \choose K} $ possible subsets, we use a randomized search strategy to find an approximate minimum to the loss.

Algorithm~\ref{alg1} first computes a random subsampling matrix as described in Section~\ref{sec:VCLDerivationComputation} above.
Here, $0<r_1\leq 1.0$ is the spatial subsampling rate, which controls the trade-off between computational complexity and accuracy.
The sampled voxels are selected within the region of interest (ROI), and for our experiments, the ROI is defined as the smallest sphere enclosing the reference object.
However, the ROI can be user-defined based on specific application requirements.
In practice, $S$ can simply be represented by a list of indices corresponding to the sampled voxels.

Next, the subsampled reference image and reconstruction bases are computed.\footnote{Additional details on the generation of the reconstruction bases for the cone-beam reconstruction problem are provided in the Supplementary Material.}
While subsampling can be applied after the reconstruction bases are computed, some reconstruction packages allow for faster back projection of sparse voxel subsets \cite{mbirjax-2024}.
In either case, the computation of these bases can be parallelized across independent processors.
Step~2 of Algorithm~\ref{alg1} then computes $(R,\gamma)$ from the subsampled bases, and Step~3 calls the \mbox{ViewSubsetSelection()} function to determine the best set of views.

Algorithm~\ref{alg2} is used to compute an optimized subset of $K$ views, $\Omega^\ast$, from the full set of views, $\Omega$.
This is done by first selecting a random subset of views, $\Omega^\ast$.
Then for each view in $\Omega^\ast$, we attempt to identify a substitute view that results in a lower VCL.
Specifically, for each view in $\Omega^\ast$, a random subset $\Omega_S$ of size $\lfloor r_2(K_{\text{max}} - K) \rfloor$ is generated from $\Omega \setminus \Omega^\ast$, where $r_2$ is the parameter that controls the trade-off of computation and thoroughness for each iteration.
In each swap, a single view from $\Omega^\ast$ is exchanged with one from $\Omega_S$.
If the swap lowers the VCL, then $\Omega^\ast$ and its corresponding loss are updated accordingly, and this is continued until no change occurs after attempting random swaps for every angle in $\Omega^\ast$.

Algorithm~\ref{alg3} evaluates the VCL.
Notice that this algorithm first selects the relevant rows and columns of $R$ and the elements of $\gamma$ to form $(R_\Omega , \gamma_\Omega)$.
This reduces computation since the $(R,\gamma)$ only need to be computed once.

\section{Experimental Results}

In this section, we compare the effectiveness of the VCLS algorithm for view selection to both conventional uniform view sampling methods and the edge projection-based view selection (EPVS) algorithm of \cite{LREdge3} for both simulated and measured X-ray CT datasets.

\subsection{Simulated Data}
\label{sec:simulated_data}

In this section, we compare VCLS to EPVS and uniform view sampling using a simulated dataset.
This allows us to compare to the exact ground truth.

Table~\ref{table:parameters} specifies the parameters used for generating and reconstructing the simulated cone-beam CT data.
Additional details on the modeling of polychromatic X-ray effects, beam-hardening correction \cite{poly_correction}, and material modeling \cite{nist_xray_mass_attenuation} are provided in the Supplementary Material.

Figure~\ref{fig:simulated_train_object}(a) shows the CAD model that was used to generate the ground truth and reference objects used in our simulations.
The ground truth object was generated by rastering a CAD model using the parameters in Table~\ref{table:parameters}.
This CAD model included extra artificial defects, such as pores and cracks embedded within the object.
The reference object is similar, but not exactly the same, as the ground-truth object.
This was done to test the robustness of our method.
More specifically, we generated the reference object by rasterizing a second CAD without defects but with a slightly different rotation.
We then rotated the reference object in 2D to visually match the position of the ground-truth object.

Figures~\ref{fig:simulated_train_object}(b) and (c) show axial slices of both the ground truth and reference objects.
A total of $720$ candidate view angles were uniformly sampled from the range of $0^\circ$ to $360^\circ$ for view selection.

Unless otherwise noted, all reconstructions used MBIR via the MBIRJAX software package \cite{mbirjax-2024}, and
all experiments were performed on 2 A100 GPUs and a 64-core CPU platform using a combination of python, numpy, and JAX software environments.
MBIRJAX automatically controls regularization using the meta-parameters of \mbox{snr\_db} and \mbox{sharpness} specified in Table~\ref{table:parameters}.

Table~\ref{table:time} compares the runtime and reconstruction loss of VCLS using different values of $r_1$ and $r_2$.\footnote{We note that Step~1 of Algorithm~\ref{alg1} used both GPUs and the sparse back-projection capability of MBIRJAX to speed computations.}
Notice that as $r_1$ and $r_2$ decrease, the computation time decreases, but the reconstruction error increases slightly.
We found $(r_1,r_2) = (0.001,0.1)$ to be a good trade-off, and used these values in all remaining experiments.

Figure~\ref{fig:recon(simulated)} shows comparisons of VCLS with uniform view sampling and the EPVS algorithm \cite{LREdge3} for the simulated cone-beam data.
The axial slices shown in Figure~\ref{fig:recon(simulated)} demonstrate that VCLS results in sharper edges and more refined texture detail than either uniform sampling or EPVS.
The plot of NRMSE versus number of views in Figure~\ref{fig:nrmse} supports this by showing that the VCLS reconstructions have lower NRMSE than the uniform or EPVS reconstructions as the number of views varies from 30 to 150.

\begin{figure*}
\centering
 \begin{subfigure}[b]{0.35\textwidth}
     \centering
     \includegraphics[width=0.8\textwidth]{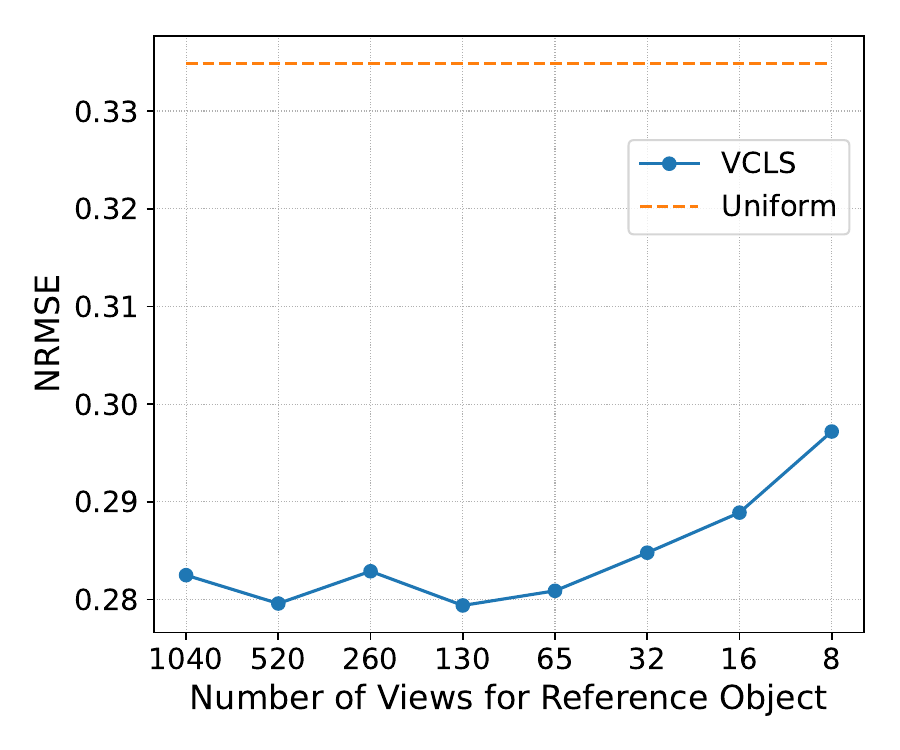}
     \caption{Reconstruction loss}
     \label{fig:nrmse_different_ref_views}
 \end{subfigure}
\begin{subfigure}[b]{0.60\textwidth}
 \centering%
 \includegraphics[width=0.8\textwidth]{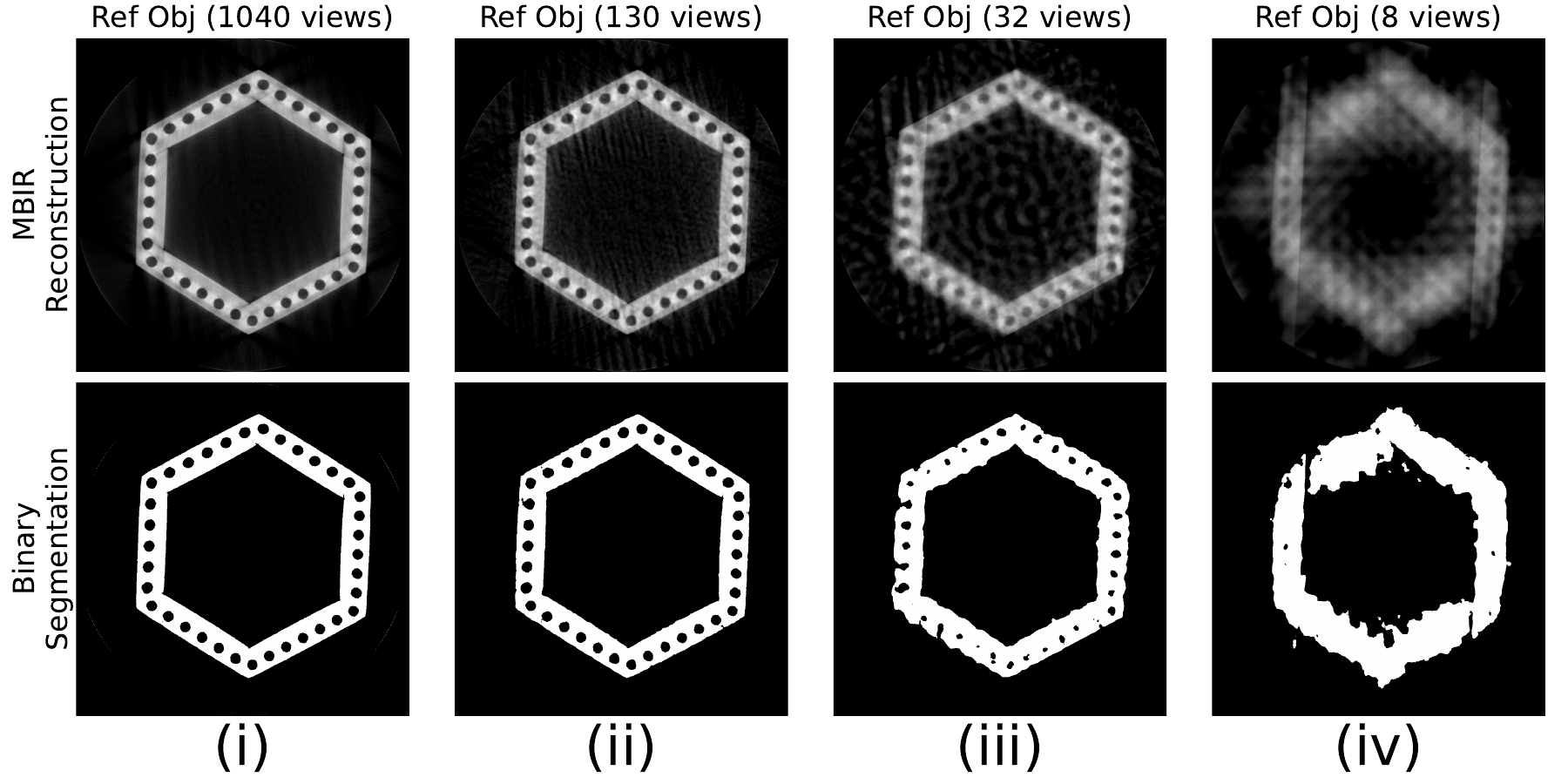}
 \caption{Reference objects}
 \label{fig:ref_comparsion}
 \end{subfigure}
\caption{
(a) Plot of reconstruction loss (NRMSE) for 40‐view reconstructions using VCLS sampling, but with varying reference object quality.
The dashed line shows baseline NRMSE with 40 uniformly sampled views.
(b) Examples of reference objects generated by reconstructing the actual object from uniformly sampled sparse views ranging from 1040 views to 8 views.
The top row shows the full axial slice of each reconstruction, and the bottom row shows the corresponding segmentation.
Note that VCLS consistently outperforms uniform sampling even when the quality of the reference object is quite poor.
}
\label{fig:ReferenceObjectQualityAssessment}
\end{figure*}

\subsection{Measured Data}
\label{sec:MeasuredData}

We next compare VCLS with EPVS and uniform view sampling using an experimentally measured axial cone-beam X-ray CT data scan of a hexagonal fuel nozzle \cite{ziabari2025xray} collected on a ZEISS Metrotom 800 system.
Table~\ref{table:measured_parameters} lists the parameters used in the experiment.
Figure~\ref{fig:LineProfile}(b) shows an axial slice of the additively manufactured hexagonal object that was used in this experiment.

In AM inspection, it is typical for a sequence of parts printed from the same CAD drawing to be scanned consecutively.
In such scenarios, the information gained from the first scanned part can be used to guide the scanning strategy for the remaining parts.
To simulate this, a binary segmentation of the reconstructed image from a dense-view scan of a separate object is used as the reference object for view selection.

Figure~\ref{fig:measured_training_object} shows the scanned object and the reference object.
Notice that the reference object does not perfectly match the scanned object, which tests the robustness of view selection under imperfect reference object conditions.

From Tuy's condition \cite{Tuy1,Tuy2}, it is known that views from $0^\circ$ to $180^\circ$ plus the cone angle (short scan) are sufficient for complete Fourier space coverage.
Therefore, in order to achieve the best results, our uniform sampling uses equally spaced samples in the range of $[0^\circ, 194.42^\circ]$, where the cone angle is $14.42^\circ$ in our setting.
However, the VCLS and EPVS algorithms are allowed to select views from the full range of $[0^\circ, 360^\circ]$ for maximum flexibility.

Unless otherwise noted, all reconstructions were performed using MBIRJAX on an A100 GPUs and a 64-core CPU platform with the parameters listed in Table~\ref{table:measured_parameters}.
Any FDK reconstructions of short scan data used Parker weighting to improve reconstruction quality and avoid corresponding short scan artifacts \cite{parker1982optimal}.

Figure~\ref{fig:MeasuredRecon} shows the reconstructed images obtained using 40 and 80 selected views, which correspond to approximately a 26$\times$ and 13$\times$ reduction in the number of measurements compared to a dense scan of 1050 views.
Notice that for both the 40 and 80 view case, the VCLS view selection produces sharper images with fewer artifacts than either uniform or EPVS view selection.
Also, VCLS achieves much better quality than FDK reconstruction with uniform view sampling shown in Figure~\ref{fig:MeasuredRecon}(i) and (v).

To demonstrate the impact of such algorithm in real world application, we considered dimensional characterization of object width and holes using sparse view reconstruction algorithms.
Such characterizations are critical in applications such as industrial metrology of parts that require accurate measurement of object dimensions.

Figure~\ref{fig:LineProfile} shows line profiles through the reconstructions along the green line shown in the dense-view reconstruction of Figure~\ref{fig:LineProfile}(b).
Then Table~\ref{table:FWHM} lists the associated measured values of the object's width using the full-width at half maximum (FWHM) measure.
The associated error is also listed for each measurement using the measured width with dense view (1050 views) as ground truth.

Notice that in Table~\ref{table:FWHM}, VCLS has the lowest measured error in both the 40 view case (0.12mm error) and in the 80 view case (0mm error).
This is consistent with the visual inspection of the line profiles for the 40 view case shown in Figure~\ref{fig:LineProfile}.
Here, the dense view result (yellow curve) is used as ground-truth.

Table~\ref{table:DiameterTable} presents a summary of an analysis of the diameters of holes in the object.
Notice that for both the 40 and 80 view case, the VCLS algorithm results in reconstructions with the lowest error in the mean diameter estimate.
Moreover, the standard deviation of the error is the lowest among the alterative view selection algorithms.
More details of the analysis are presented in the Supplementary Material.

Figure~\ref{fig:angle_selection} illustrates the view selection patterns for uniform sampling (using a short scan), EPVS, and VCLS.
Note that VCLS selects views from a narrower range of angles used by EPVS or even the short scan predicted by Tuy's condition.
This is potentially useful since it reduces the range of angles that must be obtained during a scan.
Also, notice that the VCLS views are not uniformly spaced.
Instead, there are subtle but important variations in view angle that appear to favor particularly informative views while also preserving view diversity.

\subsection{Effect of Reference Object Quality}

In this section, we investigate the effect of the reference object quality on the VCLS algorithm view selection using the measured data of Section~\ref{sec:MeasuredData}.

Figure~\ref{fig:nrmse_different_ref_views} plots the NRMSE using 40 views selected by VCLS, but based on a reference object of varying quality.
Each reference object is generated by reconstructing the object from uniformly sampled sparse views, with sampling ranging from 1040 views to only 8 views.
Figure~\ref{fig:ref_comparsion} shows the reference reconstructions used in the plot along with their corresponding segmentations.

Notice that the NRMSE of the VCLS reconstruction increases slowly, even for very low-quality reference reconstructions.
In fact, the VCLS reconstruction is still much better than the baseline (uniform view sampling) reconstruction, even when the reference is reconstructed from only 8 uniformly sampled views.
This suggests that it is possible to use even a fast ultra-sparse CT scan in order to generate a reference object of sufficient quality for use in the VCLS algorithm.

\section{Conclusion}

We present an algorithm for CT view selection which is based on the minimization of a novel view covariance loss (VCL) function.
The VCL is unique in that it is a single expression that accounts for the joint information in the views.
We show that the VCL is closely related to the Fisher information for estimating a linear projection of the object, providing a principled justification for its use.
Consequently, it encourages view selections that are both informative and diverse (i.e., uncorrelated).
The VCL also has a relatively simple closed form and can be computed efficiently for practical CT problems using straightforward acceleration methods.
Experimental results on both simulated and measured cone-beam CT data demonstrate that the view covariance loss selection (VCLS) algorithm selects views that differ subtly from uniform sampling, but that substantially improve the quality of CT reconstruction.

\section*{Acknowledgments}

This work was co-authored by UT-Battelle, LLC under contract DE-AC05-00OR22725 with the US Department of Energy (DOE) and supported by the DOE Office of Energy Efficiency and Renewable Energy (EERE), Advanced Materials \& Manufacturing Technologies Office (AMMTO).
The US government retains and the publisher, by accepting the article for publication, acknowledges that the US government retains a nonexclusive, paid-up, irrevocable, worldwide license to publish or reproduce the published form of this manuscript, or allow others to do so, for US government purposes. DOE will provide public access to these results of federally sponsored research in accordance with the DOE Public Access Plan (http://energy.gov/downloads/doe-public-access-plan)
C. Bouman was partially supported by the Showalter Trust.

\newpage
\clearpage

\bibliographystyle{IEEEtran}
\bibliography{references}

\begin{IEEEbiography}[{
\includegraphics[width=1in,height=1.25in,clip,keepaspectratio]{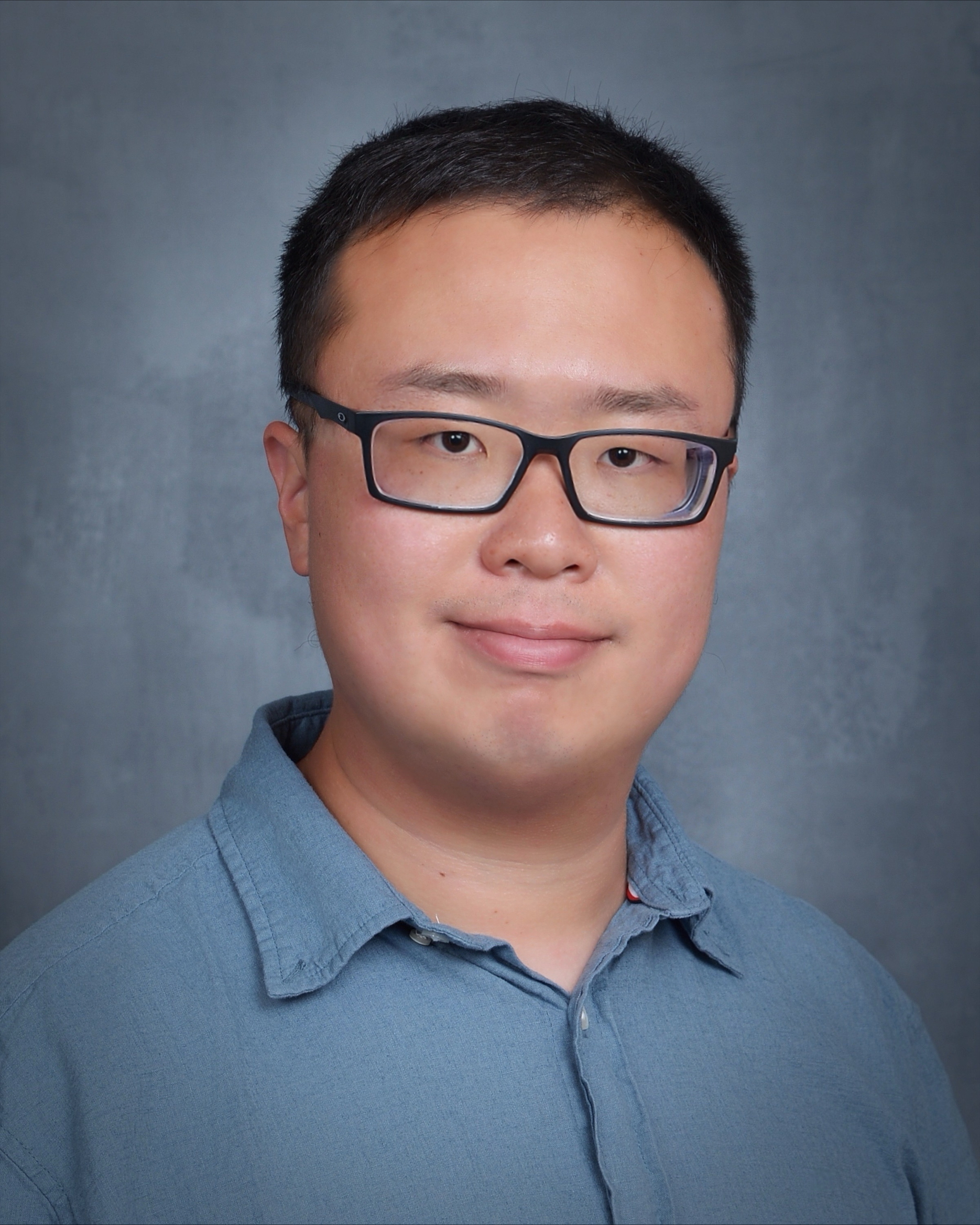}
}]{Jingsong Lin}
(Student Member, IEEE) received the B.S. degree in Electrical Engineering from both the Beijing Institute of Technology, Beijing, China, and the Australian National University, Canberra, Australia, in 2018. He received the M.S. degree in Electrical Engineering from Cornell University, Ithaca, NY, USA, in 2019. He is currently pursuing the Ph.D. degree in Electrical Engineering at Purdue University, West Lafayette, IN, USA. His research interests include computational imaging, inverse problems, and signal processing.
\end{IEEEbiography}

\begin{IEEEbiography}[{
\includegraphics[width=1in,height=1.25in,clip,keepaspectratio]{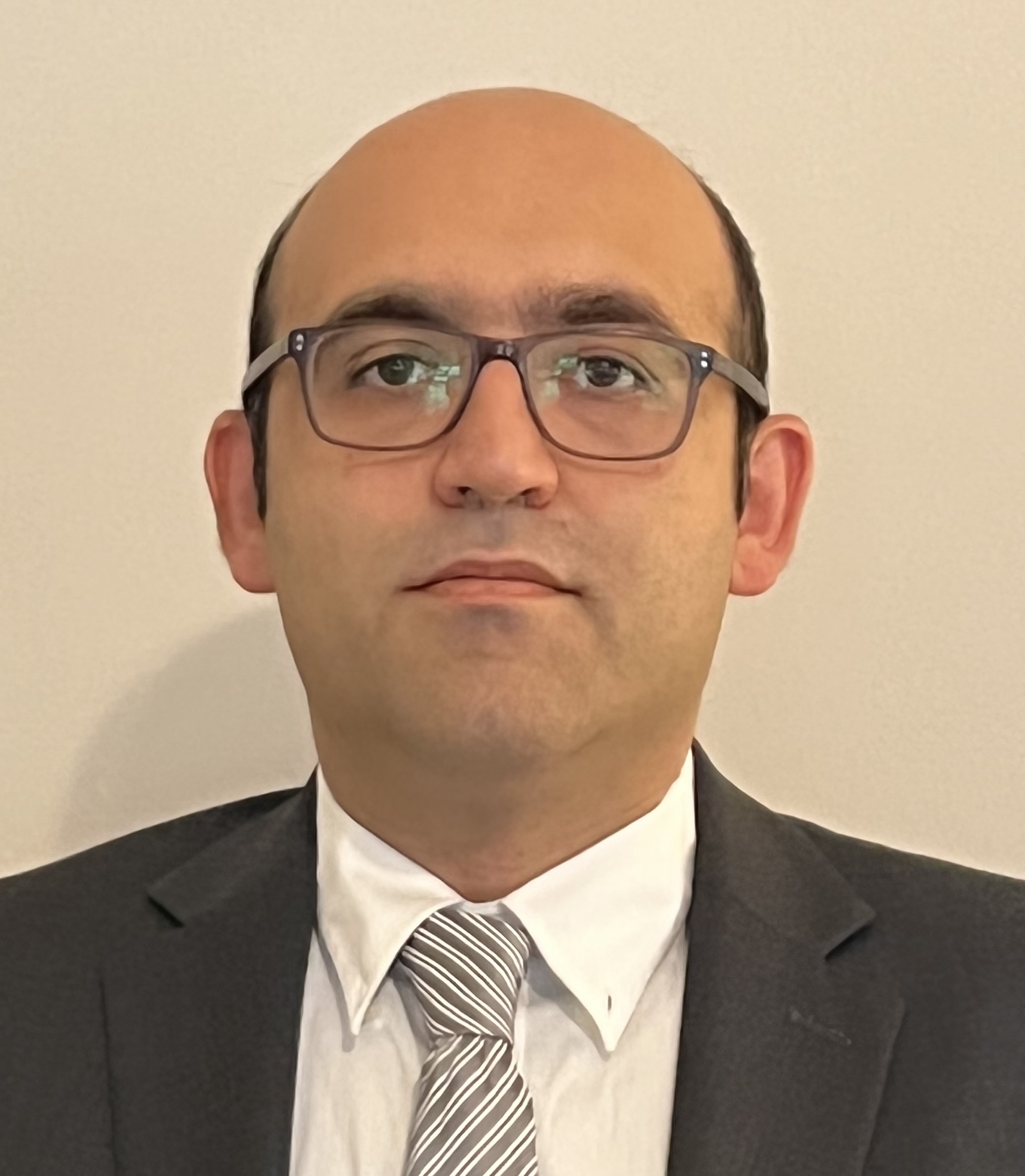}
}]{Amirkoushyar Ziabari}
(Senior Member, IEEE) is a senior R\&D staff scientist in the Multimodal Sensor Analytics group at the Electrification and Energy Infrastructure Division (EEID) at Oak Ridge National Laboratory (ORNL). Before joining ORNL, Dr. Ziabari was a Postdoc in the Integrated Imaging group at the Department of Electrical and Computer Engineering (ECE) at Purdue University. He also received his PhD from the department of ECE at Purdue University in August 2016. His research interests have been at the intersection of physics, signal processing and machine/deep learning, where he has used image/signal processing and machine learning (ML) ideas to explore multiscale experimental physics phenomena. He is currently working on data analytic, data driven and physics-based image reconstruction and segmentation for advanced manufacturing (AM) in order to improve non-destructive evaluation (NDE) and the state-of-the-art in real time monitoring of the AM process. Dr. Ziabari is also leading several other projects on developing ML and computational imaging algorithms for image analysis and reconstruction in scientific imaging, including electron microscopy, electron tomography, neutron CT, coded-aperture imaging, etc. Dr. Ziabari is a Senior Member of the IEEE and a member of the IEEE Computational Imaging Technical Committee.
\end{IEEEbiography}

\begin{IEEEbiography}
[{
\includegraphics[width=1in,height=1.25in,clip,keepaspectratio]{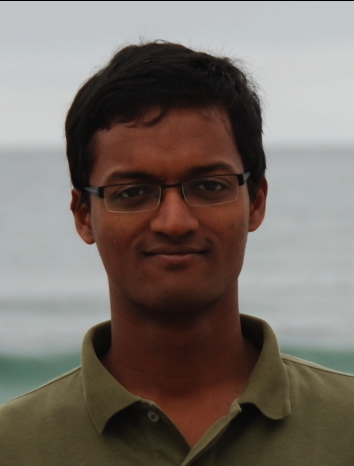}
}]
{Singanallur V. Venkatakrishnan}
(Senior Member, IEEE) received the Ph.D. degree from the School of Electrical and Computer Engineering, Purdue University, West Lafayette, IN, USA, in 2014. He is currently a Senior R\&D Staff Member with Multimodal Sensor
Analytics Group, Oak Ridge National Laboratory developing computational imaging algorithms in support of the lab’s efforts in ultrasound, X-ray, electron and neutron-based systems. His research interests include computational imaging, inverse problems, and machine learning. Dr. Venkatakrishnan is a Senior Area Editor of IEEE Transaction On Computational Imaging.
\end{IEEEbiography}

\begin{IEEEbiography}[{
\includegraphics[width=1in,height=1.25in,clip,keepaspectratio]{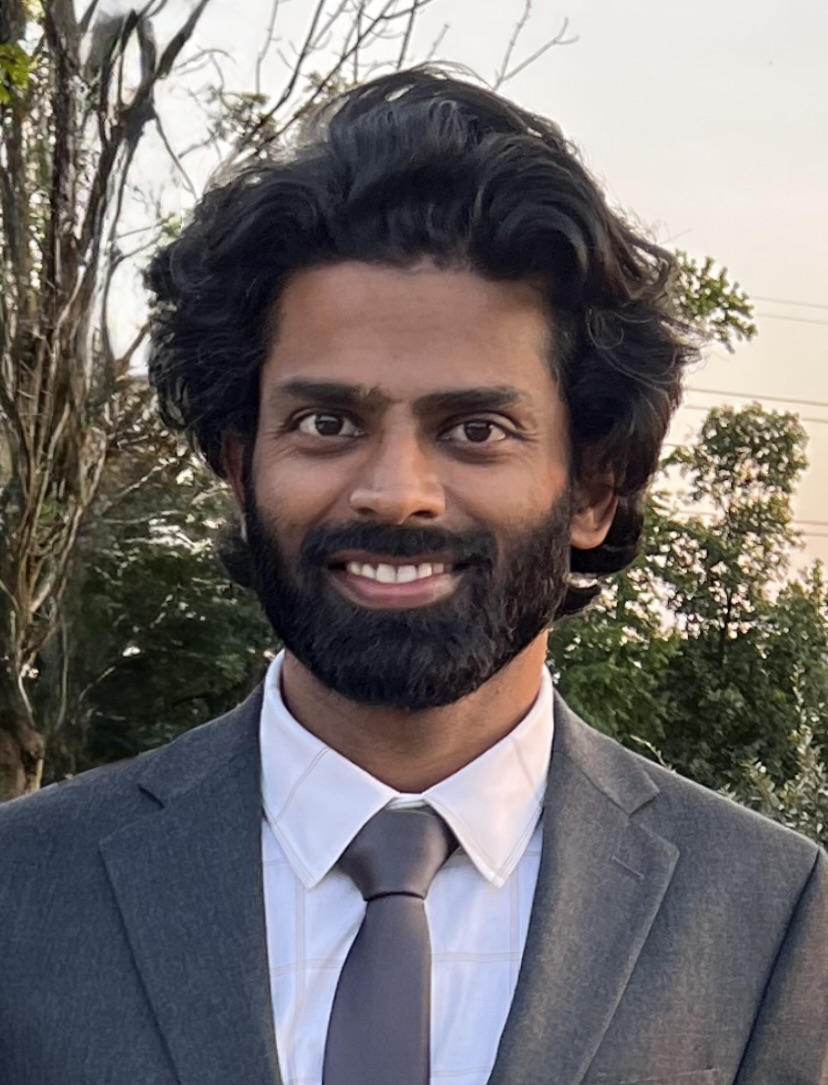}
}]{Obaidullah Rahman}
 is a Data Analytics Researcher at Oak Ridge National Laboratory, specializing in computational imaging, X-ray computed tomography (CT), and electron microscopy.
 He earned his Ph.D. and M.S. in Electrical Engineering from the University of Notre Dame, and a B.Tech. in Electronics Engineering from Aligarh Muslim University.
 His research spans deep learning algorithms for CT reconstruction, defect characterization in additive manufacturing, and characterization of fuel cell microscopy samples. Obaidullah has done internship at General Electric Healthcare where he worked on low-dose medical CT images and developed analytical and deep learning solutions to combat low-SNR artifacts.
\end{IEEEbiography}

\begin{IEEEbiography}[{
\includegraphics[width=1in,height=1.25in,clip,keepaspectratio]{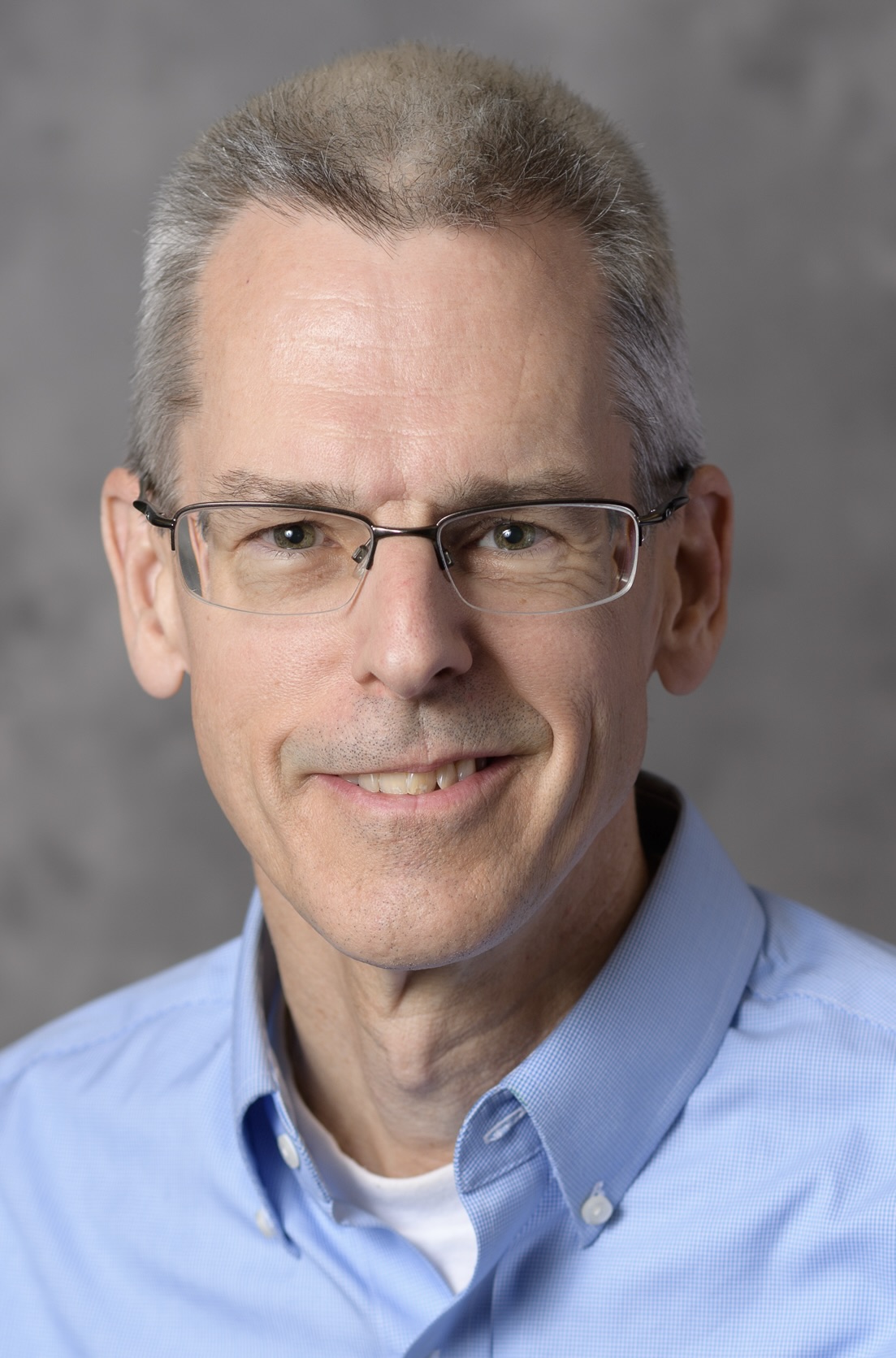}
}]{Gregery T. Buzzard}
(Senior Member, IEEE) received the Ph.D. degree from the University of Michigan, Ann Arbor, MI, USA.
He was a Postdoctoral with Indiana University, Bloomington, IN, USA, and Cornell University, Ithaca, NY, USA.
He is currently a Professor of mathematics and the Director of the Center for Computational and Applied Mathematics with Purdue University, West Lafayette, IN, USA, where he was the Mathematics Department Head for seven years.
His research has led to theoretical advances in dynamical systems and experiment design and to new algorithms for image and volume reconstruction for a variety of sensing modalities.
The unifying ideas in his recent work are iterative methods for image reconstruction and reduction of uncertainty through appropriate measurement schemes.
\end{IEEEbiography}

\begin{IEEEbiography}[{
\includegraphics[width=1in,height=1.25in,clip,keepaspectratio]{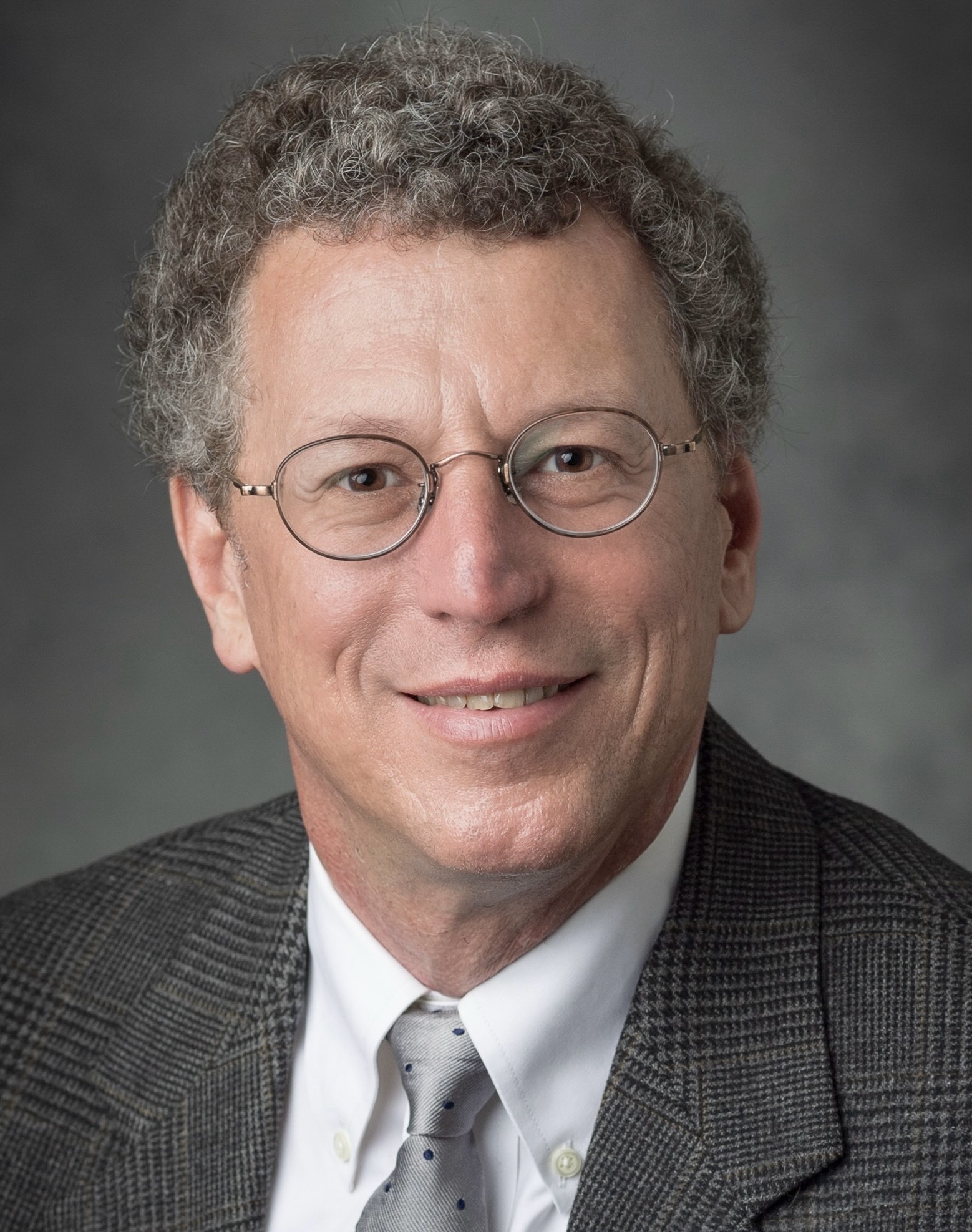}
}]{Charles A. Bouman}
(Life Fellow, IEEE) is the Showalter Professor of Electrical and Computer Engineering and Biomedical Engineering at Purdue University. He received his B.S.E.E. from the University of Pennsylvania, M.S. from the University of California at Berkeley, and Ph.D. from Princeton University in 1989.
Prof. Bouman’s research in computational imaging is focused on the integration of statistical signal processing, physics, and computation to solve problems with applications in healthcare, scientific, industrial, and consumer imaging. He is a member of the National Academy of Inventors, a Life Fellow of the IEEE, a Fellow and Honorary Member of the IS\&T, and a Fellow of SPIE and AIMBE. He is currently a Vice President of the IS\&T. From 2000 to 2004, he served as the Vice President of Publications for the IS\&T. In 2022 he was made an Honorary Member of the IS\&T, and in 2023, he was given the IS\&T Service Award. He received the 2021 IEEE Signal Processing Society, Claude Shannon-Harry Nyquist Technical Achievement Award, the 2014 Electronic Imaging Scientist of the Year award, and in 2020, his paper on Plug-and-Play Priors won the SIAM Imaging Science Best Paper Prize. In 2003, he founded the IS\&T Computational Imaging Conference and led the creation of the IEEE Transactions on Computational Imaging.
\end{IEEEbiography}

\end{document}


\AddToShipoutPictureFG*{\AtPageUpperLeft{\put(\LenToUnit{0.6in},\LenToUnit{-0.42in}){\parbox[t]{\dimexpr\paperwidth-1.2in\relax}{\fontsize{6.5}{7.5}\selectfont\centering \copyright~2025 IEEE. Personal use of this material is permitted. Permission from IEEE must be obtained for all other uses, in any current or future media, including reprinting/republishing this material for advertising or promotional purposes, creating new collective works, for resale or redistribution to servers or lists, or reuse of any copyrighted component of this work in other works. Accepted version of the article published in IEEE Transactions on Pattern Analysis and Machine Intelligence, vol.~48, no.~10, pp.~11335--11345, Oct.~2026. DOI: 10.1109/TPAMI.2025.3600072}}}}

\maketitle

The following contains supplementary material for the submitted manuscript.

\section{Derivation of Fisher Information Results}

Let the forward model be
$$
y = T^t ( \tilde{x} + \epsilon ) \ ,
$$
where $\epsilon\sim N(0,I)$.
So, $y \sim N( \mu (\tilde{x}) , \Sigma (\tilde{x}) )$ where
\begin{align*}
\mu(\tilde{x} ) &= T^t \tilde{x} \\
\Sigma(\tilde{x}) &= T^t T = R \ ,
\end{align*}
Then the Fisher information for the estimation of $\tilde{x}$ given $y$ is given by
$$
{\cal I}_{\tilde{x} }= T^t \Sigma^{-1} T = T^t R^{-1} T \ .
$$

If we define a linear function $\tilde{x} = c \, \eta$ where $\eta \in \mathbb{R}$ and $c \in \mathbb{R}^N$, then the Fisher information for the estimation of $\eta$ given $y$ is given by
$$
{\cal I}_{\eta} =
\frac{\partial \tilde{x}}{\partial \eta}^t {\cal I}_{\tilde{x}}
\frac{\partial \tilde{x}}{\partial \eta}
=
c^t {\cal I}_{\tilde{x}} c
= c^t T R^{-1} T^t c \
$$
Then if we choose $c=x/\|x\|$ where $x$ is the reference object, we get the following expression for the Fisher information
$$
{\cal I}_{VCL} = \gamma^t R^{-1} \gamma = 1 - VCL \ .
$$

\begin{figure*}
\centering
\includegraphics[width=0.9\textwidth]{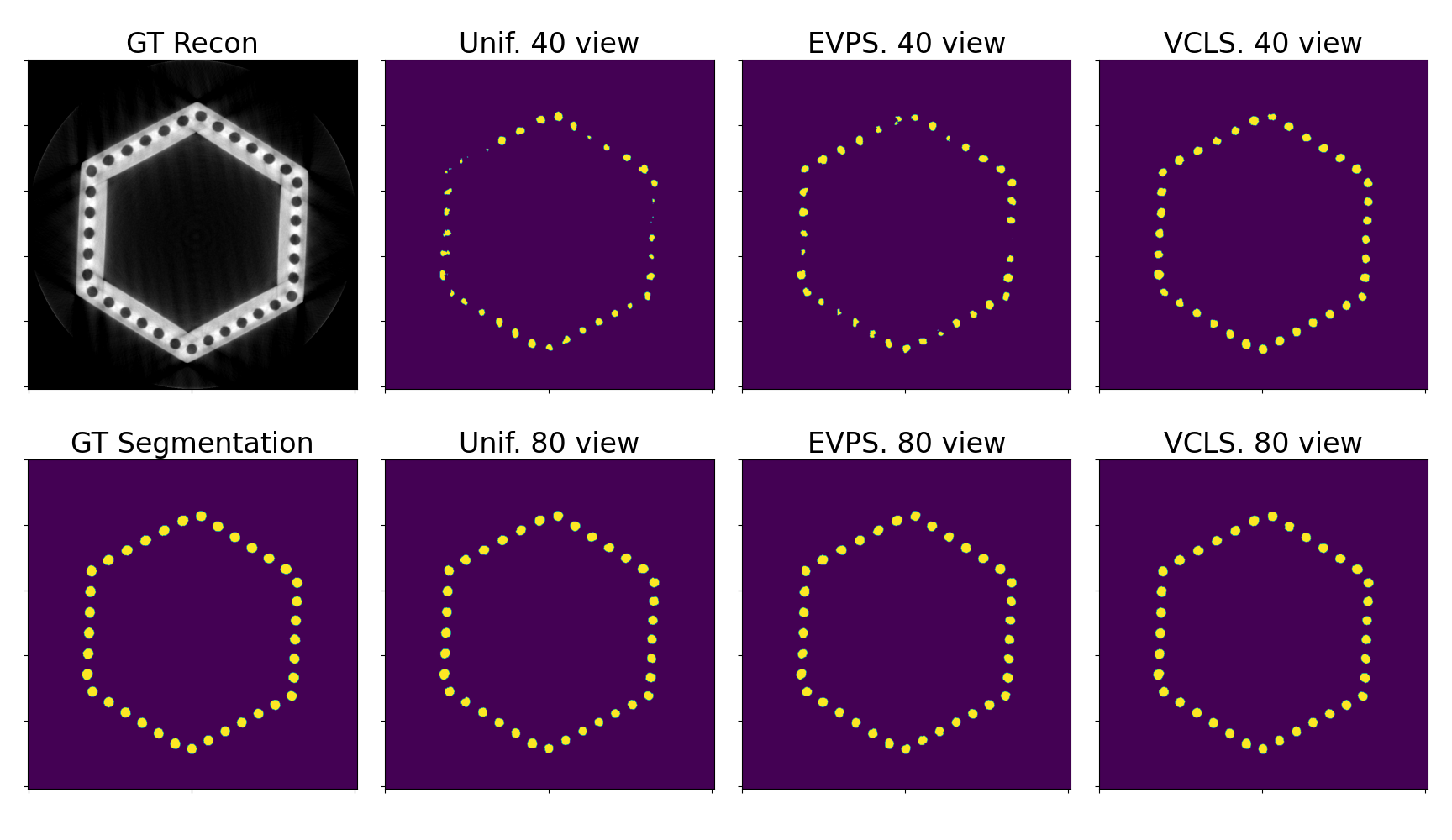}
\caption{Comparison of the segmentation of holes in hexagonal fuel nozzle using different view selection algorithms for 40 and 80 view cases.}
\label{fig:hole_segment}
\end{figure*}

\begin{figure}
\centering
\includegraphics[width=0.4\textwidth]{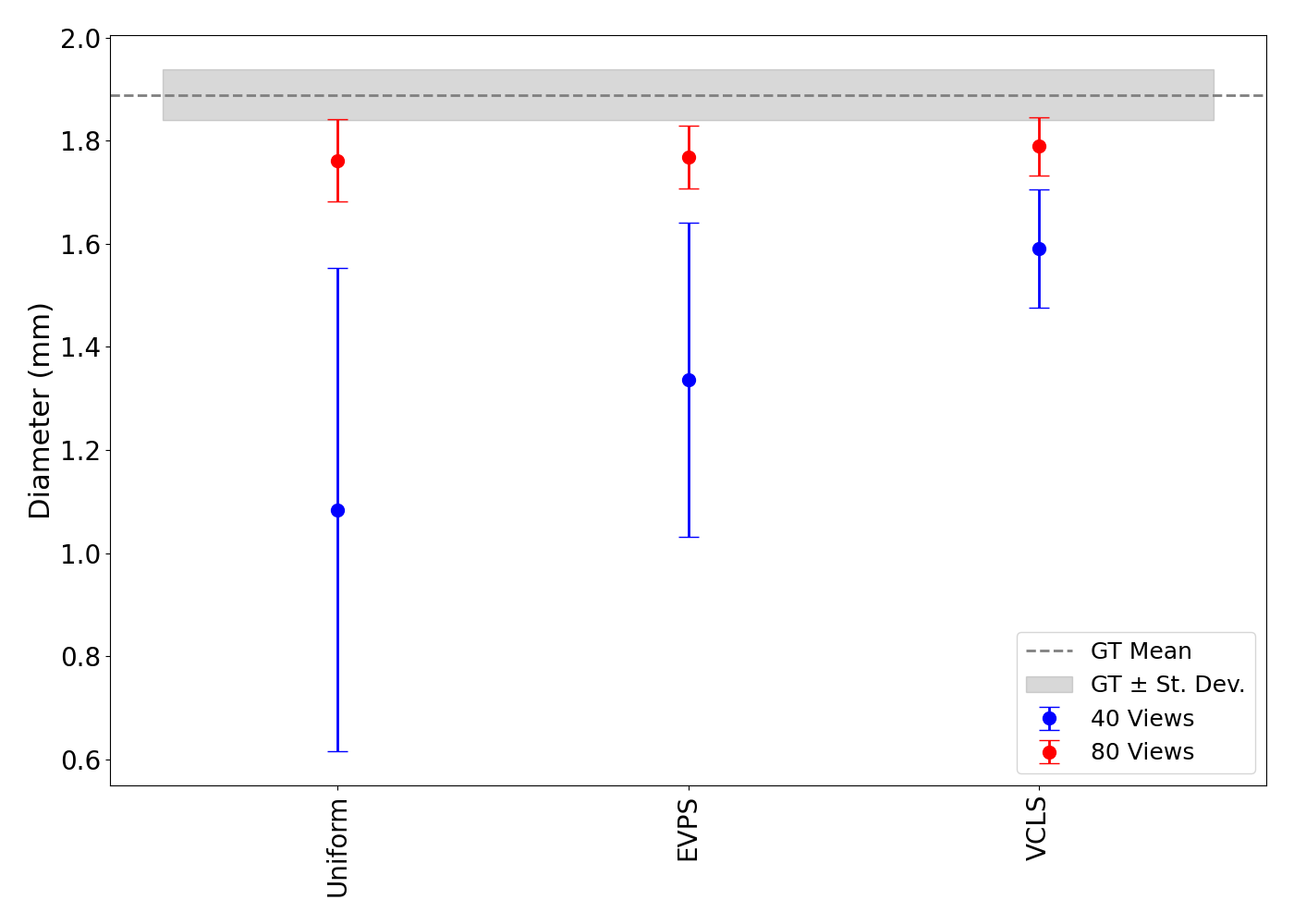}
\caption{
Statistical comparison of measured hole diameters in hexagonal fuel nozzle using different view selection algorithms.
Note that VCLS produces a diameter mean closer to the ground truth and a smaller standard deviation compared to uniform sampling and EPVS in both the 40-view and 80-view cases.
}
\label{fig:hole_dist}
\end{figure}

\section{Generation of Reconstruction Bases}
\label{sec:GenerationFDKFiltered Back-Projections}

This section provides implementation details for generating the reconstruction bases in cone-beam CT using Equation 2 in the main paper.
We compute reconstruction bases using MBIRJAX, but similar methods can be used to compute the bases with any cone-beam tomography package that provides forward and backward projection functions.

First, we generate the simulated projection views $y_k = A_k x$ where $x$ denotes the reference object and $A_k$ is the distance driven cone-beam projector at rotation angle $\theta_k$ as described in \cite{balke2018separable}.

Let $y_k (m,n)$ denote the $k^{th}$ view where $m$ is the detector row and $n$ is the detector channel.
Here we assume that the axis of rotation of the object is aligned with the channel of the detector.
Let ``iso'' denote a point in space formed by the projection of the point source location onto the line formed by axis of rotation.
Then let ``detector iso'' denote the projection of iso onto the 2D plane of the detector.
And let $u(n)$ denote the distance in cm along the channel direction from the \mbox{detector iso} to the $n^{th}$ column of the detector, and let $v(m)$ denote the distance in cm along the row direction from the \mbox{detector iso} to the $m^{th}$ row of the detector.

Then using these definitions, define the weighting function
$$
w(m,n) = \frac{ D_{sd}}{\sqrt{D_{sd}^2 + (u(n))^2 + (v(m))^2} } \ ,
$$
where $D_{sd}$ is the source to detector distance in cm, and define the discrete-time impulse response
$$
h(n) = \frac{1}{2} \mbox{sinc} (n) - \frac{1}{4} \mbox{sinc}^2 (n) \ .
$$

Then for FDK, the filtered projection is given by
$$
g_k (m,n) = \left[ w(m,n) y_k (m,n) \right] * h(n) \ ,
$$
where $*$ denotes 1D discrete-time convolution along the channel direction.

For FBP, the filtered projection is given by
$$
g_k (m,n) = y_k (m,n) * h(n) \
$$
where $*$ again denotes 1D discrete-time convolution along the channel direction.

For both FDK and FBP, the reconstruction basis functions are then formed by back projecting the filtered projection.
$$
T_k = A_k^t g_k \ .
$$
Notice that $T_k$ does not include proper scaling for FDK or FBP reconstruction.
However, this is fine since, as noted in Equation 2 in the main paper, the VCL function is invariant to arbitrary scaling by a constant $\alpha$.

\section{Generation of Simulated Data}
\label{sec:GenerationSimulatedData}

This section explains the procedure for generating the simulated projection data as described in Section 5.1 in the main paper, and how a polynomial correction function is used to reduce the beam-hardening effects.

For each view angle, synthetic projections are generated from the ground truth using a polychromatic X-ray source model to better simulate the beam-hardening effects.
The synthetic cone-beam projection data, $y$, is computed as
\begin{equation*}
    y = - \log \dfrac{I}{I_o}
\end{equation*}
where $I$ is the object scan intensity and $I_o$ is the blank scan intensity.

Then $I$ and $I_o$ are computed as follows.
First, the the projection path lengths through the object are computed as
$$
y_l = Ax \ ,
$$
where $A$ as units of \mbox{cm} and $x$ is a binary representation of the object.
So then each entry of $y_l$ has units of \mbox{cm} and represents the total integrated path length of the projection through the object.

From this, we next compute the expected measurement of the object scan given by
$$
\bar{I} = \tau_o \sum_{m=0}^{M-1} f(E_m)  e^{-\rho \mu(E_m)y_l} (E_m - E_{m-1}) \ ,
$$
where the sum approximates an integral; $E_m$ are discrete energy levels used for the Riemann approximation in units of \mbox{keV}; \(f(E)\) is the simulated effective X-ray power spectrum generated with SpekPy \cite{spekpy} in units of $\mbox{photons}/(\mbox{sec}-\mbox{keV})$; $\tau_o$ is the time window of the measurement in units of \mbox{sec}; and $\rho$ and $\mu(E)$ are the density and mass attenuation coefficient for 316L stainless steel obtained from NIST \cite{nist_xray_mass_attenuation} in units of $\mbox{gm}/\mbox{cm}^{3}$ and $\mbox{cm}^2/\mbox{gm}$, respectively.

From this, we generate the random object scan measurement given by
$$
I \sim \mbox{Pois}(\bar{I}) \ ,
$$
where $\mbox{Pois}(\bar{I})$ generates a vector of independent Poisson distributed random variables with mean $\bar{I}$

The blank scan measurement is simulated by the deterministic quantity given by
$$
I_o = \tau_o \sum_{m=0}^{M-1} f(E_m) (E_m - E_{m-1}) \ .
$$
We model the blank scan as deterministic because, in practice, it is measured with sufficient duration that the photon-counting noise is negligible compared to the object scan.

Reconstruction using the projections data, $y$, suffers from beam-hardening effects.
One method to mitigate this is to apply a polynomial correction function to the projection data \cite{poly_correction}.
In our experiments, the polynomial correction function used is
\begin{equation*}
y' = 0.58y + 0.52y^2 + 0.30y^3 - 0.09y^4 + 0.007y^5 \ .
\end{equation*}

\begin{table}
\caption{EPVS Parameter Settings}
\centering
\renewcommand{\arraystretch}{1.2}
\setlength{\tabcolsep}{3pt}
\begin{tabular}{|c|c|c|c|}
\hline
\textbf{Param.} &\textbf{Function} & \makecell[tc]{\bf Simulated \\ \bf Data} & \makecell[tc]{\bf Measured \\ \bf Data} \\
\hline
$\lambda$ & \makecell{view information versus \\ diversity weighting} & $1.0$ & $0.5$ \\
\hline
$\beta$ & \makecell{edge alignment parameter} & $75.0$ & $50.0$ \\
\hline
$\gamma$ & \makecell{pairwise view \\ dispersion parameter} & $1.0$ & $1.0$ \\
\hline
$r$ & \makecell{Reference object\\subsampling rate} & $5$ & $5$ \\
\hline
\end{tabular}
\label{table:epvs_parameter}
\end{table}

\section{Metrological Analysis of Measured Data}
\label{sec:metrological_analysis}

In this section, we provide further metrological analysis of the fuel nozzle data measured on the ZEISS Metrotom 800 system.
More specifically, we segment out the holes in the object, and we quantify the accuracy of the segmented holes relative to ground truth.

Figure~\ref{fig:hole_segment} shows the resulting binary segmentations of the holes for ground truth and each of the view selection algorithms using both 40 and 80 views.
Visual inspection indicates that for 40 views, VCLS achieves more accurate segmentation of the holes.

Figure~\ref{fig:hole_dist} presents a quantitative analysis of the segmented hole diameters.
Notice that VCLS achieves a mean diameter closer to the ground truth for both 40 and 80 views than EVPS or uniform view sampling.
Also, the standard deviation of the VCLS estimates is lower than that for EVPS or uniform view sampling for both the 40 and 80 view cases.

\section{Fourier Interpretation of VCLS }

\begin{figure}
  \centering
  \begin{subfigure}[b]{0.22\textwidth}
    \centering
    \includegraphics[width=0.9\textwidth]{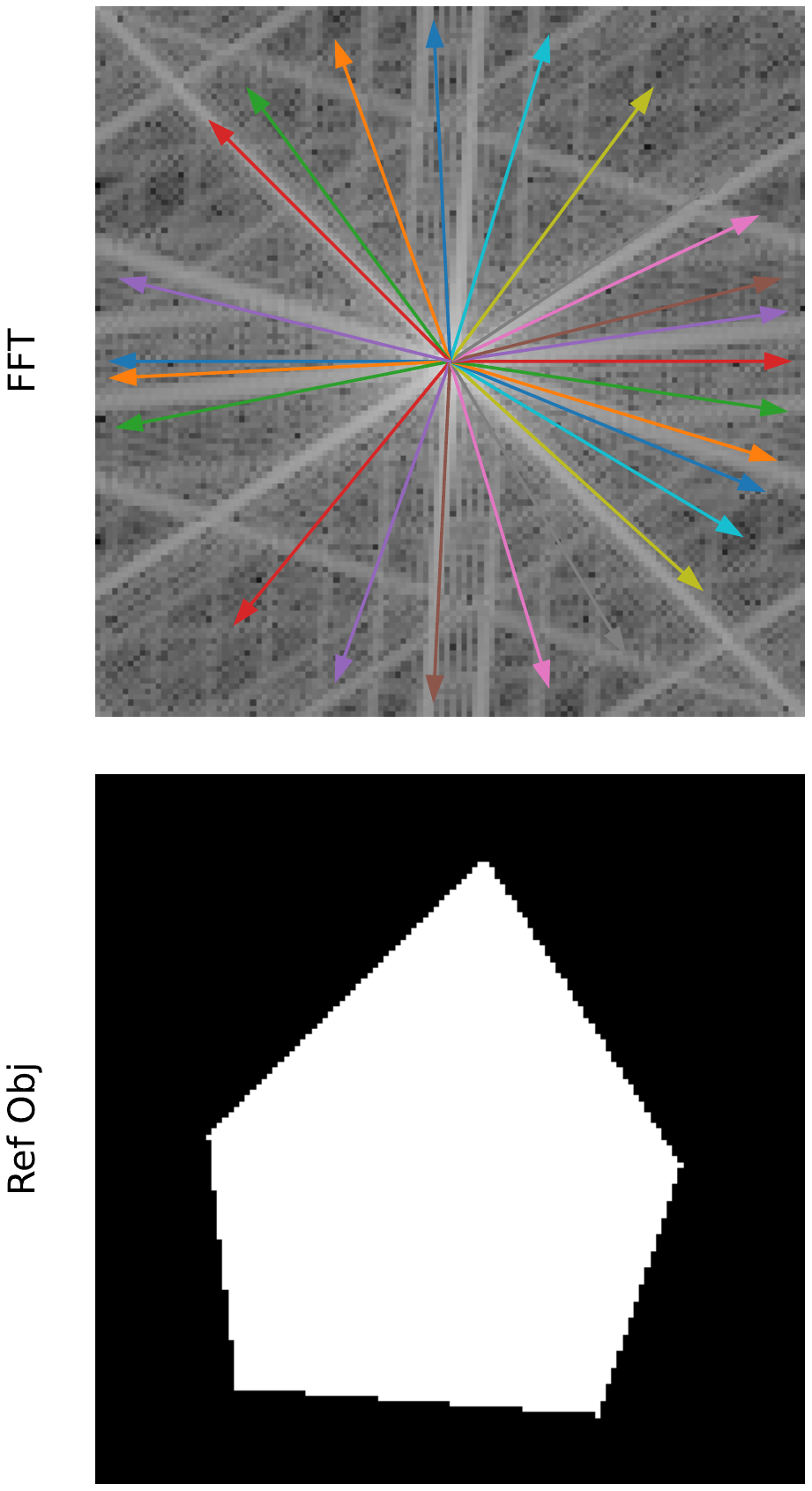}
    \caption{Polygon-shape object}
  \end{subfigure}
  \begin{subfigure}[b]{0.22\textwidth}
    \centering
    \includegraphics[width=0.9\textwidth]{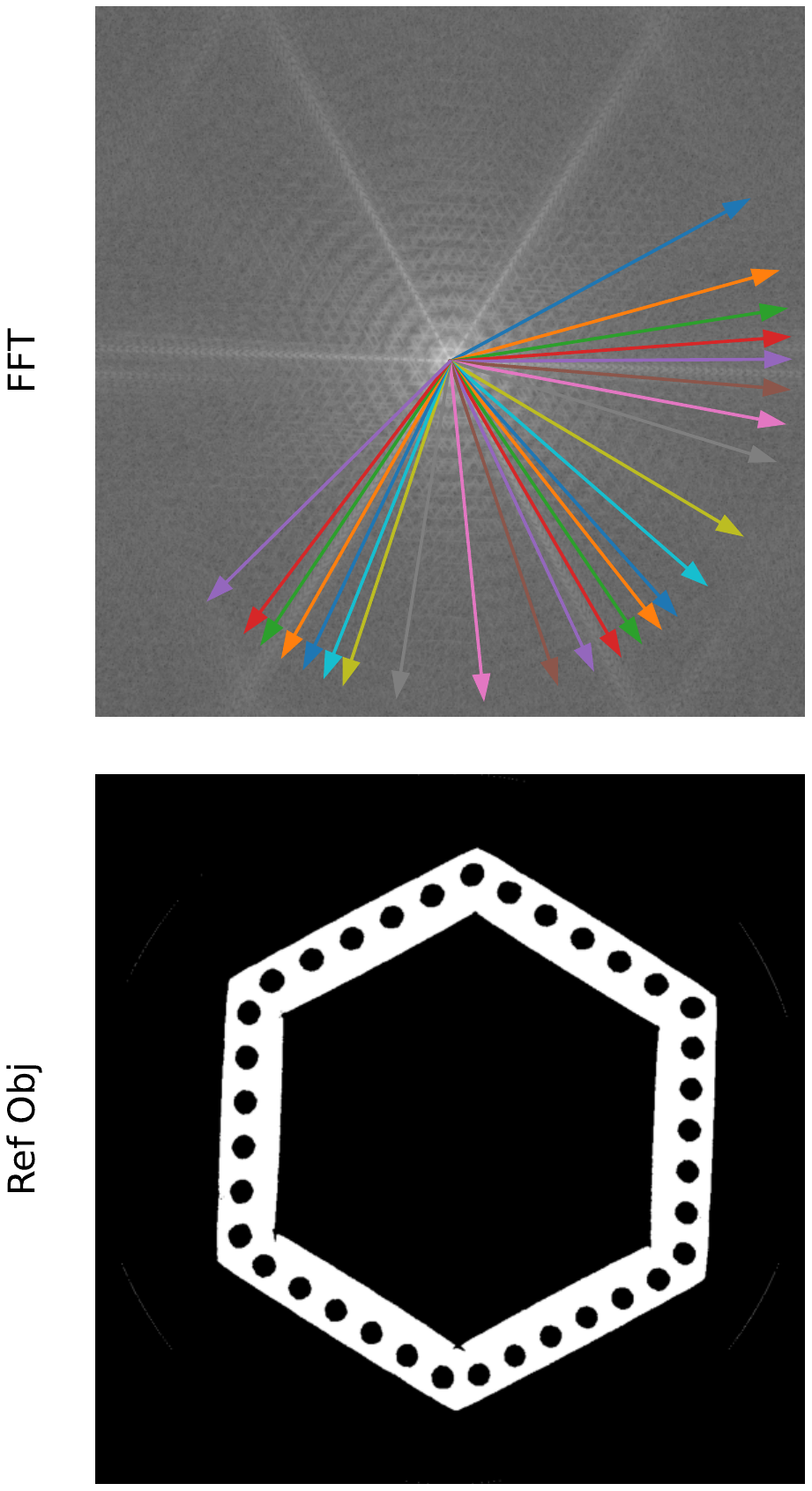}
    \caption{Hexagonal fuel nozzle}
  \end{subfigure}
  \hfill
  \caption{Plot of the Fourier spectrum of the reference objects overlaid with the 25 views selected by VCLS.}
  \label{fig:fft_plot}
\end{figure}

In this section, we examine the relationship between the Fourier transforms of the reference objects and the corresponding view angles selected by VCLS.
Figure~\ref{fig:fft_plot} presents the $\log_{10}$ magnitude of the FFT of the two reference objects overlaid with the 25 views selected by VCLS.
The first object is a simple polygon and the second is the hexagonal fuel nozzle used in previous results.
Note that the view angles were rotated by $90\deg$ in order to align with the Fourier features.

Notice that while there is some alignment between the view angles and the Fourier features, it is not as great as one might expect.
This may be at least partially accounted for by the fact that, for a cone-beam projector, each view contains projects over a range of angles.
Therefore the best view angle for each edge depends not only on its angle, but also its offset from the center of rotation.

\section{Implementation of EPVS}
\label{sec:ImplementationEPVS}

EPVS \cite{LREdge3} is an adaptive view selection algorithm designed for large-scale cone-beam CT problems, and it serves as one of the baseline methods in our experiments.

EPVS works by sequentially selecting the next view angle, $\theta^\ast$, given the previous set of view angles, $\Omega^\ast$, using the recursion
\begin{align*}
\theta^\ast =  \arg \max_{\theta \in \Omega} \big\{
& \mbox{EdgeAlignment}(\theta; x, \beta ) \\
&+ \lambda\, \mbox{Dispersion}(\theta, \Omega^\ast; x, \gamma) \big\}
\end{align*}
where $\Omega$ is the full set of possible views under consideration, $x$ is the reference image, $\Omega^\ast$ is the previously selected view angles, $\beta$ and $\gamma$ are parameters, and $\lambda$ is a weighting that controls the importance of view diversity as compared to the amount of information individually provided by the view.

Table~\ref{table:epvs_parameter} lists the parameters we used in EPVS for both the simulated and measured data.
The algorithm parameters were adjusted to yield the best reconstruction quality for the experimental data presented.

\bibliographystyle{IEEEtran}
\bibliography{references}